\documentclass[twocolumn]{aastex631}
\usepackage{amsmath,amssymb}
\usepackage{float}
\usepackage{hyperref}
\usepackage{dsfont}

\shorttitle{Too many quenched galaxies at $z>3$}
\shortauthors{Zhang et al.}

\graphicspath{{./}{figures/}}

\begin{document}

\title{RUBIES spectroscopically confirms the high number density of quiescent galaxies from $\mathbf{2<z<5}$}

\author[0000-0001-6454-1699]{Yunchong Zhang} 
\affiliation{Department of Physics and Astronomy and PITT PACC, University of Pittsburgh, Pittsburgh, PA 15260, USA}

\author[0000-0002-2380-9801]{Anna de Graaff}
\affiliation{Max-Planck-Institut f\"ur Astronomie, K\"onigstuhl 17, D-69117 Heidelberg, Germany}

\author[0000-0003-4075-7393]{David J. Setton} 
\thanks{Brinson Prize Fellow} \affiliation{Department of Astrophysical Sciences, Princeton University, 4 Ivy Lane, Princeton, NJ 08544, USA}

\author[0000-0002-0108-4176]{Sedona H. Price}
\affiliation{Space Telescope Science Institute, 3700 San Martin Drive, Baltimore, Maryland 21218, USA}
\affiliation{Department of Physics and Astronomy and PITT PACC, University of Pittsburgh, Pittsburgh, PA 15260, USA}

\author[0000-0001-5063-8254]{Rachel Bezanson}
\affiliation{Department of Physics and Astronomy and PITT PACC, University of Pittsburgh, Pittsburgh, PA 15260, USA}

\author[0000-0003-3021-8564]{Claudia del P. Lagos}
\affiliation{International Centre for Radio Astronomy Research (ICRAR), M468, University of Western Australia, 35 Stirling Hwy, Crawley, WA 6009, Australia}

\author[0000-0002-7031-2865]{Sam E. Cutler}
\affiliation{Department of Astronomy, University of Massachusetts, Amherst, MA 01003, USA}

\author[0000-0002-2446-8770]{Ian McConachie}
\affiliation{Department of Astronomy, University of Wisconsin-Madison, 475 N. Charter St., Madison, WI 53706 USA}

\author[0000-0001-7151-009X]{Nikko J. Cleri}
\affiliation{Department of Astronomy \& Astrophysics, The Pennsylvania State University, University Park, PA 16802, USA} 
\affiliation{Institute for Computational \& Data Sciences, The Pennsylvania State University, University Park, PA 16802, USA}
\affiliation{Institute for Gravitation and the Cosmos, The Pennsylvania State University, University Park, PA 16802, USA}

\author[0000-0003-3881-1397]{Olivia R. Cooper}
\affiliation{Department of Astronomy, The University of Texas at Austin, Austin, TX, USA}
\affiliation{Cosmic Dawn Center (DAWN), Copenhagen, Denmark}

\author[0000-0003-0205-9826]{Rashmi Gottumukkala}
\affiliation{Cosmic Dawn Center (DAWN), Copenhagen, Denmark}
\affiliation{Niels Bohr Institute, University of Copenhagen, Jagtvej 128, Copenhagen, Denmark}

\author[0000-0002-5612-3427]{Jenny E. Greene}
\affiliation{Department of Astrophysical Sciences, Princeton University, 4 Ivy Lane, Princeton, NJ 08544, USA}
\author[0000-0002-3301-3321]{Michaela Hirschmann}
\affiliation{Institute of Physics, Laboratory for Galaxy Evolution, EPFL, Observatory of Sauverny, Chemin Pegasi 51, CH-1290 Versoix, Switzerland}

\author[0000-0002-3475-7648]{Gourav Khullar}
\affiliation{Department of Astronomy, University of Washington, Physics-Astronomy Building, Box 351580, Seattle, WA 98195-1700, USA}
\affiliation{eScience Institute, University of Washington, Physics-Astronomy Building, Box 351580, Seattle, WA 98195-1700, USA}

\author[0000-0002-2057-5376]{Ivo Labbe}
\affiliation{Centre for Astrophysics and Supercomputing, Swinburne University of Technology, Melbourne, VIC 3122, Australia}

\author[0000-0001-6755-1315]{Joel Leja}
\affiliation{Department of Astronomy \& Astrophysics, The Pennsylvania State University, University Park, PA 16802, USA} 
\affiliation{Institute for Computational \& Data Sciences, The Pennsylvania State University, University Park, PA 16802, USA}
\affiliation{Institute for Gravitation and the Cosmos, The Pennsylvania State University, University Park, PA 16802, USA}

\author[0000-0003-0695-4414]{Michael V.\ Maseda}
\affiliation{Department of Astronomy, University of Wisconsin-Madison, 475 N. Charter St., Madison, WI 53706 USA}

\author[0000-0003-2871-127X]{Jorryt Matthee}
\affiliation{Institute of Science and Technology Austria (ISTA), Am Campus 1, 3400 Klosterneuburg, Austria}

\author[0000-0001-8367-6265]{Tim B. Miller}
\affiliation{Center for Interdisciplinary Exploration and Research in Astrophysics (CIERA), Northwestern University, 1800 Sherman Ave, Evanston, IL 60201, USA}

\author[0000-0003-2804-0648]{Themiya Nanayakkara}
\affiliation{Centre for Astrophysics and Supercomputing, Swinburne University of Technology, Melbourne, VIC 3122, Australia}

\author[0000-0002-1714-1905]{Katherine A. Suess}
\affiliation{Department for Astrophysical \& Planetary Science, University of Colorado, Boulder, CO 80309, USA}

\author[0000-0001-9269-5046]{Bingjie Wang}
\affiliation{Department of Astronomy \& Astrophysics, The Pennsylvania State University, University Park, PA 16802, USA} 
\affiliation{Institute for Computational \& Data Sciences, The Pennsylvania State University, University Park, PA 16802, USA}
\affiliation{Institute for Gravitation and the Cosmos, The Pennsylvania State University, University Park, PA 16802, USA}

\author[0000-0001-7160-3632]{Katherine E. Whitaker}
\affiliation{Department of Astronomy, University of Massachusetts, Amherst, MA 01003, USA} 
\affiliation{Cosmic Dawn Center (DAWN), Copenhagen, Denmark}

\author[0000-0003-2919-7495]{Christina C. Williams}
\affiliation{NSF National Optical-Infrared Astronomy Research Laboratory, 950 North Cherry Avenue, Tucson, AZ 85719, USA}

\begin{abstract} 

We present the number density of massive ($ \mathrm{ log (M_{*}/M_{\odot}) > 10.3} $) quiescent galaxies at $2<z<5$ using JWST NIRSpec PRISM spectra. This work relies on spectra from RUBIES, which provides excellent data quality and an unparalleled, well-defined targeting strategy to robustly infer physical properties and number densities. We identify quiescent galaxy candidates within RUBIES through principal component analysis and construct a final sample using star formation histories derived from spectro-photometric fitting of the NIRSpec PRISM spectra and NIRCam photometry. By inverting the RUBIES selection function, we correct for survey incompleteness and calculate the number density of massive quiescent galaxies at these redshifts, providing the most complete spectroscopic estimates prior to cosmic noon to date. We find that early massive quiescent galaxies are surprisingly common ($\gtrsim 10^{-5}$ Mpc$^{-3}$ by $4<z<5$), which is consistent with previous studies based on JWST photometry alone and/or in smaller survey areas. We compare our number densities with predictions from six state-of-the-art cosmological galaxy formation simulations. At $z>3$, most simulations fail to produce enough massive quiescent galaxies, suggesting the treatment of feedback and/or the channels for early efficient formation are incomplete in most galaxy evolution models.

\end{abstract}


\section{Introduction} \label{sec:intro}

How the most massive galaxies stopped forming stars (``quenched") remains one of the most important but mysterious questions in galaxy evolution. Evidence suggests that massive quiescent galaxies ($\mathrm{M_{*}>10^{11}M_{\odot}}$) are already in place in the first $\mathrm{\sim 1\,Gyr}$ after the Big Bang \citep[e.g.,][]{Carnall.etal.2024,Weibel.etal.2024,deGraaff.etal.2025}. The efficient formation channel for these early galaxies is heavily debated \citep[e.g.,][]{Liu.etal.2022,Dekel.etal.2023,Ferrara.etal.2023}. Certain mechanisms must be triggered in these quiescent systems to rapidly shut down their previously efficient star formation, despite the fact that cosmic star formation density is rising in this epoch \citep[e.g.,][]{Madau.etal.2014}. Common speculations of these mechanisms include feedback from compact star formation \citep[e.g.,][]{Diamond-Stanic.etal.2012,Sell.etal.2014} or from active galactic nuclei (AGN; e.g., \citealp{Croton.etal.2006,Fabian.etal.2012}). These processes are capable of exhausting the central cool gas reservoir that fuels the star formation in galaxies, by either removing them mechanically through outflow, heating them to disrupt gravitational collapse, or preventing further gas accretion from the cosmic environment \citep[e.g.,][]{Bezanson.etal.2019,Whitaker.etal.2021,Williams.etal.2021,D'Eugenio.etal.2024,Belli.etal.2024}. Historically, these mechanisms are incorporated into simulations to recreate the observed galaxy populations \citep[e.g.,][]{Croton.etal.2006,Bower.etal.2006,Sijacki.etal.2007,Lagos.etal.2008,Somerville.etal.2008}.

Since the launch of the JWST \citep{Gardner.etal.2006,Gardner.etal.2023}, tremendous progress has been made in selecting candidates of high redshift ($z>3$) massive quiescent galaxies \citep[e.g.,][]{Carnall.etal.2023,Valentino.etal.2023} as well as in confirming them spectroscopically \citep[e.g.,][]{Carnall.etal.2024,Nanayakkara.etal.2024,deGraaff.etal.2025}. As these quenched objects are dominated by older stellar populations of high mass-to-light ratios, we can mitigate systematic bias in stellar mass estimation due to outshining from young ($\mathrm{< 100\,Myr}$ old) stellar populations \citep[e.g.,][]{Gimenez-Arteaga.etal.2023,Papovich.etal.2023}. Moreover, the number densities of these massive quiescent systems in the early universe can place valuable constraints on our models of galaxy evolution \citep[e.g.,][]{DeLucia.etal.2024,Lagos.etal.2024}, which are mostly calibrated to observations at later times \citep[e.g.,][]{Somerville&Dave2015}. Constrained by the age of the universe when these early massive quiescent galaxies are observed, their inferred star formation histories (SFHs) implies high or even close-to-theoretical-limit \citep{Carnall.etal.2024,Glazebrook.etal.2024,deGraaff.etal.2025,Turner.etal.2025} baryon conversion efficiency in the early dark matter halos, implying extreme formation efficiency. Alternatively, studies have explored how modifying inference assumptions — such as the initial mass function (IMF; e.g., \citealp{vanDokkum.etal.2024}) and abundance patterns \citep[e.g.,][]{Park.etal.2024b,Beverage.etal.2025} — can mitigate the inferred extreme stellar masses and reconstructed SFHs.

Several studies have attempted to probe the number densities of the quiescent population at the earliest possible time, using photometrically selected candidates from the early JWST imaging fields \citep[e.g.,][]{Carnall.etal.2023,Valentino.etal.2023,Alberts.etal.2024,Long.etal.2024,Baker.etal.2025}. However, the classical photometric rest-frame color selection techniques utilized successfully at $z<2$ \citep[e.g.,][]{Williams.etal.2009,Whitaker.etal.2011} face various issues at $z>3$. Firstly, photometry alone is not sufficient to break the age-dust degeneracy, which compromises the purity of color-selected quiescent samples, especially when the wavelength coverage is limited to NIRCam filters ($\lambda<4.4\mu$m) \citep[e.g.,][]{Antwi-Danso.etal.2023}. Secondly, early color selections miss the youngest (mean stellar age $\mathrm{< 300\,Myr}$) quiescent galaxies, which become more common at $z>3$ \citep[e.g.,][]{Belli.etal.2019,Baker.etal.2024,Park.etal.2024}. Expanded color selections have been motivated to include younger quiescent galaxies \citep[e.g.,][]{Belli.etal.2019,Baker.etal.2024}. Nevertheless, the ranges of these new photometric criteria and the available wavelength sampling create a patchwork of poorly understood selection effects and interloper populations, which could be driving the high variance in abundances measured between different surveys. Finally, photometric rest-frame colors will always be sensitive to systematic uncertainties due to photometric redshifts, which depend on the nature of the photometry. 

Deep near-infrared (NIR) spectroscopy can more robustly confirm the quiescent nature of these high-redshift candidates. But most studies thus far have been limited to single objects or targeted follow-up of small samples \citep[e.g.,][]{Carnall.etal.2023b,Carnall.etal.2024,Glazebrook.etal.2024,Onoue.etal.2024,Weibel.etal.2024,deGraaff.etal.2025}. To date, only a few sizable spectroscopic samples have attempted to constrain the number density of quiescent galaxies, finding largely consistent statistics as photometric studies. \cite{Baker.etal.2024} assembled a spectroscopic sample by combining multiple early JWST observing programs. However, fully characterizing the selection function of such a sample is challenging due to the wide range of selection criteria that saw these sources placed on slits. Spectroscopic follow-up can be used to assess the purity of photometric samples; \cite{Nanayakkara.etal.2025} estimated $\sim 80\%$ purity of their photometric parent sample, which could then be used to update initial number density estimates \citep{Schreiber.etal.2018a}. However, given that the photometric sample utilizes the classical rest-frame UVJ color selection, this is likely an underestimate at $z>3$ \citep[e.g.,][]{Antwi-Danso.etal.2023,Long.etal.2024}. \citet{Park.etal.2024} attempted a less direct but complementary method and measured the number density of quiescent galaxies at $z>3$ by reconstructing the formation histories of a spectroscopic sample of quiescent galaxies observed at lower redshifts $z\sim2$ \citep{Park.etal.2024}. However, this approach is sensitive to the choice of model assumptions \citep{Carnall.etal.2019,Leja.etal.2019,Suess.etal.2022,Park.etal.2024b} and will likely underestimate the true number densities due to progenitor bias. Finally, all early estimates of high redshift galaxy number densities are limited by the relatively small field of view of JWST. Therefore, cosmic variance sets a floor on the precision of this measurement (e.g., $\sim20\%$, for a combined area of $\mathrm{145\,arcmin^2}$; \citealp{Valentino.etal.2023}).

These early studies of quiescent galaxy number densities at face value indicate an overabundance of quiescent galaxies at $z>3$ compared to predicted values from six state-of-the-art cosmological galaxy formation simulations \citep[e.g.,][]{Lagos.etal.2025}. Given that tension exists relative to all models despite the wide variation in the implementation of the AGN feedback needed to quench massive galaxies, additional model development may be needed. However, given the known limitations in the observational measurements (e.g., due to sample purity, contamination, or cosmic variance), we must also demonstrate that this discrepancy is not caused by imprecise observational results. Thus, as we make this comparison between observed and predicted number densities, all observational studies must strive to identify quiescent galaxies that are fully representative of the full population over the largest possible area. 

In this paper, we present a spectroscopic census of massive quiescent galaxies at $2<z<5$ over a cosmologically relevant volume. This work relies on data collected from the JWST RUBIES (Red Unknowns: Bright Infrared Extragalactic Survey; GO\#4233, PIs: A. de Graaff and G. Brammer; \citealp{deGraaff.etal.2024rubies}) Program. RUBIES is an NIRSpec \citep{Boker.etal.2023} micro shutter array (MSA) survey that spans color space and aims to provide a census of the reddest high-redshift sources. The well-characterized selection function (in empirical color-magnitude space) and high completeness for rare high-redshift objects of RUBIES allow us to robustly estimate the quiescent galaxy number density, despite the rare nature of the population. We identify quiescent galaxy candidates within the RUBIES dataset from their rest-frame optical to NIR spectral energy distributions (SEDs) using principal component analysis (PCA; similar to \citealt{Wild.etal.2014}. We then perform spectro-photometric stellar population synthesis modeling to finalize a robust sample of massive quiescent galaxies. 

The paper is organized as follows. In Section \ref{Sec: data}, we describe the spectroscopic and imaging data used in our analysis. We detail the identification of quiescent galaxies, including PCA analysis and spectro-photometric modeling, in Section \ref{Sec: Methods}. We compare our selection to conventional rest-frame color selection methods and account for targeting selection and incompleteness in Section \ref{Sec: Results}. In Section \ref{Sec: Number Density}, we present the number density of massive quiescent galaxies in our sample, which we compare to literature results from both observations and simulations. We summarize our findings in Section \ref{Sec: Summary}. Throughout this paper, we assume a flat $\mathrm{\Lambda CDM}$ cosmology with $ \mathrm{\Omega_{\Lambda} = 0.69}$, $ \mathrm{\Omega_{m} = 0.31}$, and $\mathrm{H_{0} = 67.66 \, km\,s^{-1}\,Mpc^{-1}}$ as reported in \cite{Planck18}. We adopt a Chabrier IMF \citep{Chabrier.etal.2003}.

\section{Data} \label{Sec: data}
\subsection{RUBIES NIRSpec/MSA Spectroscopy}
 The JWST/NIRSpec Spectroscopy used in this work is taken from the RUBIES program (GO\#4233, PIs: A. de Graaff and G. Brammer; \citealp{deGraaff.etal.2024rubies}). RUBIES targets two extragalactic legacy fields: the Extended Groth Strip (EGS) and Ultra-deep Survey (UDS), which are parts of the Cosmic Assembly Near-infrared Deep Extragalactic Legacy Survey (CANDELS; \citealp{Grogin.etal.2011,Koekemoer.etal.2011}). The parent photometric catalog from which the RUBIES targets were selected was created using public JWST/NIRCam mosaic imaging of EGS from the Cosmic Evolution Early Release Science Survey (CEERS; \citealp{Bagley.etal.2023,Finkelstein.etal.2023,Finkelstein.etal.2025}) and in UDS from the Public Release IMaging for Extragalactic Research (PRIMER; \citealp{Donnan.etal.2024}). RUBIES prioritizes sources that are red in color ($\mathrm{F150W-F444W > 2}$), bright ($\mathrm{F444W<27}$), or have high photometric redshifts ($\mathrm{z_{phot} >6.5}$), which were computed with \texttt{EAZY} \citep{Brammer.etal.2008}. The parent photometric catalog of RUBIES was derived by merging public catalogs that are available on the DAWN JWST Archive (DJA) and was visually inspected to remove invalid entries due to artifacts or pipeline errors. This catalog is used for the completeness correction described in Section \ref{Sec: Correcting for Incompleteness}.

All RUBIES spectra used in this work were reduced with  \texttt{msaexp}\footnote{\href{https://github.com/gbrammer/msaexp}{https://github.com/gbrammer/msaexp}} \citep{Brammer.etal.2023b}, corresponding to version 3 of NIRSpec data (described in detail in \citealt{deGraaff.etal.2024rubies}) released on DJA\footnote{\href{https://s3.amazonaws.com/msaexp-nirspec/extractions/nirspec_rubies_graded_v3.html}{https://s3.amazonaws.com/msaexp-nirspec/extractions/nirspec\_rubies\_graded\_v3.html}} \citep{deGraaff.etal.2024,Heintz.etal.2025}. In brief, uncalibrated exposures were taken from the Mikulski Archive for Space Telescopes (MAST) \footnote{All the JWST data used in this paper can be found in MAST: \dataset[10.17909/sjsj-8p46]{http://dx.doi.org/10.17909/sjsj-8p46}.} and processed through the Detector1Pipeline steps of the standard JWST pipeline, where a mask was inserted for large cosmic-ray snowball events \citep{Rigby.etal.2023} calculated with \texttt{snowblind} \citep{Davies.etal.2024} and a $1/f$ correction was applied. After individual slits are identified, the 2D unrectified spectra are flat-fielded and flux-calibrated. Generally, RUBIES spectra are reduced with two different background subtraction strategies. However, not all RUBIES spectra have robust spectroscopic redshift quality (grade 3 as defined in \citealt{deGraaff.etal.2024rubies}) in both reduction versions. In one version, the sky background is removed ``locally'' by taking image differences of the 2D spectra taken at the three spacecraft nod offset positions. Additionally, a second version of RUBIES spectrum reduction was implemented with a global sky subtraction, where a global background is obtained by interpolating spectra from empty slits over the survey footprint and then subtracted from each source spectrum given its spatial location. The global background subtraction version is recommended for bright extended sources. The massive quiescent galaxies that this project aims to search for are bright and expected to be resolved. For any RUBIES spectra included in this analysis (see Section \ref{sec:pca_selection} for selection criteria), we adopt the global subtraction where available (535 sources), defaulting to local background subtraction for eight sources. Once background-subtracted, 1D spectra were optimally extracted \citep{Horne1986} from the rectified 2D spectra (detailed description is available in \citealp{Heintz.etal.2025}). Finally, using the a priori source position within the shutter and assuming an azimuthally symmetric Gaussian profile, an effective extended-source path-loss correction for light outside the slitlet for each source is derived and applied to the spectrum.

Although this work focuses on the NIRSpec/PRISM spectra as they provide homogeneous coverage of spectral information that constrains the stellar population modeling at $2<z<5$, we note that RUBIES also collected NIRSpec G395M spectra ($2.9-5.3 \mu m$; $R\sim1000$) for these sources. The inclusion of G395M spectra in stellar population synthesis modeling is left for future studies.

\subsection{NIRCam photometry in the EGS and UDS} \label{Sec: photometry}

The RUBIES photometric targeting catalog was not measured from point spread function (PSF)-matched NIRCam images and is therefore not optimal for further scientific analysis. For further analysis, we utilize public PSF-matched photometry catalogs in EGS and UDS\footnote{The CEERS/EGS photometry catalog \citep{Wright.etal.2024} can be accessed at\href{https://zenodo.org/records/11658282}{https://zenodo.org/records/11658282} \citep{zenodo.data}. The PRIMER/UDS photometry catalog \citep{Cutler.etal.2024} was created with a similar methodology but is not yet publicly released.}. These catalogs were created with NIRCam F115W, F150W, F200W, F277W, F356W, F410M, and F444W mosaic images of EGS and UDS fields from CEERS and PRIMER (for UDS/PRIMER, the catalog also includes F090W) that are reduced with \texttt{grizli} \citep{Brammer.etal.2023a}, corresponding to version 7.2 on DJA \citep{Valentino.etal.2023}. These mosaic images have pixel scales of $0.04''$/pixel. All sources were first detected from a long-wavelength sky-subtracted noise-equalized F277W+F356W+F444W stack image, using \texttt{SEP} \citep{Barbary.etal.2016}. Then, PSFs were empirically built in each band, following \cite{Skelton.etal.2014} and \cite{Whitaker.etal.2019}, and PSF-matching kernels were produced using \texttt{Pypher} \citep{Boucaud.etal.2016} and convolved to degrade all NIRCam images to F444W resolution. Finally, aperture photometry was extracted using circular apertures with \texttt{SEP} \citep{Bertin1996,Barbary.etal.2016}. These aperture fluxes were corrected to total based on each object's circularized Kron radius and PSF size. The detailed description of photometric catalog construction can be found in \cite{Weaver.etal.2024}. We note that the public versions of the PSF-matched catalogs do not explicitly include rest-frame colors, which we calculate from the \texttt{eazy-py} files using the \texttt{fsps} templates.  

The light distributions of all candidate sources exceed the aperture size of JWST MSA micro shutters. Any potential color gradients can lead to an intrinsic mismatch between the color information obtained from spectroscopy and the standard circular aperture photometry. The spectra mostly target the galaxy cores, while the aperture photometry can include a larger fraction of the full source. To mitigate this potential systematic, we extract a second set of photometry inspired by \cite{Nanayakkara.etal.2025}. This customized photometry approximates the rectangular aperture of the central MSA micro shutter and is measured from the same PSF-matched $40\, mas$ EGS/CEERS or UDS/PRIMER mosaics. This conservative choice is made because the spectroscopic optimal extraction mostly weights light gathered in the central micro shutter. Therefore, we assume that this photometry will be sufficiently close to the region probed by the MSA spectroscopy. 

The customized NIRCam photometry further anchors the color information of these systems during modeling when used jointly with the NIRSpec PRISM spectra, which already have outstanding flux calibration. However, the derived luminosity-dependent physical properties, such as stellar mass and star-formation rate, would be underestimated because the slit-like aperture cannot capture the total light of these extended objects. Therefore, we take the NIRCam F444W photometry documented in the existing public catalogs (produced with the methodology detailed in \citealp{Weaver.etal.2024}) and apply corrections to any spectro-photometric modeling results that depend on the normalization of the SED. We defer the details of such corrections to Section \ref{sec:SED fitting}. For a given source, the public catalogs provide a series of photometry measurements extracted from apertures of various sizes. The catalog also provides a ``best-use” value extracted from the recommended aperture size, which varies from source to source depending on the morphology. We default to the photometry from the recommended aperture (i.e., values reported in the \texttt{SUPER} catalog) for the total F444W photometry with which we derive the scaling factor for our models. For sources in this sample, these apertures are typically $0.70"$ or $1.00"$ in diameter.

\section{Method} \label{Sec: Methods}
\subsection{PCA Selection}\label{sec:pca_selection}

Although two sets of rest-frame colors (e.g., U-V and V-J) have been used to identify Balmer/$\mathrm{4000\AA}$ breaks in quiescent galaxies, the full SEDs probed by the $\mathrm{0.6-5.3\mu m}$ NIRSpec PRISM contain significantly more information. Historically, PCA has been applied to both spectroscopic and photometric data to classify SEDs of extragalactic sources \citep[e.g.,][]{Wild.etal.2014}. The PCA primarily identifies the main characteristics shared among the dataset, which are known as the eigenspectra. For each galaxy, PCA returns a set of super colors (SC), whose total number equals the number of eigenspectra. These super colors are the normalizing coefficients of eigenspectra in the reconstruction of the original spectrum: 
\begin{equation}
    F_{\lambda, reconstructed}(\lambda) = F_{\lambda,mean}(\lambda) + \sum_{i=0}^{n-1} SC_{i}*F_{\lambda,eigen,i}(\lambda) ,
\end{equation}
where $\rm F_{\lambda, reconstructed}(\lambda)$ is the reconstructed spectrum, $\rm F_{\lambda, mean}(\lambda)$ is the mean spectrum, $\rm F_{\lambda,eigen,i}(\lambda)$ are the eigen spectra, and $n$ is the total number of super colors or eigenspectra. As $n$ increases, the reconstructed spectrum converges to the original spectrum. Once a number $n$ is chosen such that the residual between the reconstructed spectrum and the original spectrum is negligible, each set of $n$ super colors can be taken as the low-dimensional representation of each spectrum.

Compared with traditional color selection, PCA enables visualization and classification of observed SEDs without relying on successful fits of spectral synthesis models. At $z>3$, early JWST surveys have uncovered novel extragalactic sources that are difficult to constrain by traditional spectral synthesis models, such as ``Little Red Dots" (LRD) \citep[e.g.,][]{Greene.etal.2024,Matthee.etal.2024}. In addition, the classical color selection of quiescent galaxies at this epoch is shown to be incomplete and impure \citep[e.g.,][]{Belli.etal.2019,Antwi-Danso.etal.2023,Baker.etal.2024,Park.etal.2024}. Despite efforts to modify these color criteria, their selection effects are yet poorly understood. PCA provides a data-driven and model-independent approach for selecting and studying the properties of these sources. In this section, we explore the classification of RUBIES SEDs at $2<z<5$ with PCA. Using PCA classification, we then focus on investigating the pre-selection of quiescent galaxies that enables the cost-efficient identification of this population.

Although we limit our identification of quiescent galaxies to the RUBIES dataset, for which the targeting strategy is well-known, we define the eigenspectra and super color space by leveraging the full set of public NIRSpec/PRISM spectra from the DJA as follows. For RUBIES spectra, we include all spectra that correspond to sources with $\mathrm{m_{F444W} <24}$. This ensures a complete selection below this magnitude limit in RUBIES. We further include any PRISM spectra logged with a signal-to-noise ratio $\mathrm{(SNR) \geq 5}$ on DJA from other JWST spectroscopic programs (including but not limited to JADES \citealp{D'Eugenio.etal.2025}; CEERS \citealp{Finkelstein.etal.2023}; UNCOVER \citealp{Price.etal.2025}; GTO WIDE \citealp{Maseda.etal.2024})\footnote{The spectra used to construct the eigenspectra are drawn from the following programs: GTO-1180, 1181, 1210, 1211, 1213, 1214, 1215, 1286; ERS-1345 (CEERS) PI: Finkelstein; 1433 PI: Coe, 1747 PI: Roberts-Borsani; 2028 PI: Wang; 2073 PI: Hannawi; 2198 PI: Barrufet; 2282 PI: Coe; 2651 (UNCOVER) PIs: Labbe and Bezanson; 2565 PI: Glazebrook 2750; 2750 PI: Arrabal Haro; 2756 PI: Chen; 2767 PI: Kelly; 3073 PI: Castellano; 3215 PI: Eisenstein; 4233 (RUBIES) PI: de Graaff; 4446 PI: Frye; 4557 PI: Yan, 6368 (CAPERS) PI: Dickinson; 6541 PI: Egami; 6585 PI: Coulter}. All selected spectra have reduction version 3 on DJA and robust spectroscopic redshifts. At low resolution ($R \sim 100$ for the NIRSpec PRISM), the selection of quiescent galaxies relies heavily on the information from the Balmer/$\mathrm{4000 \AA}$ break as well as the UV slope. Given the wavelength coverage of the NIRSpec/PRISM ($0.6 - 5.3 \mu m$), we adopt a lower redshift cut of $z \geq 2$. 

We de-redshift and resample these spectra onto a universal rest-frame wavelength grid, using a package implemented in Python \texttt{SpectRes} \citep{Carnall.etal.2017}. We exclude spectra with significant data discontinuities between $\mathrm{2500 \AA < \lambda_{rest-frame} < 7400 \AA}$ due to reduction quality issues, resulting in a final number of 3206 PRISM spectra. To minimize the contamination of strong emission line flux in our continuum-based selection, we mask out rest-frame $\mathrm{4775 - 5125 \AA} $ ($\mathrm {[OIII]+H\beta}$) and $\mathrm{6375 - 6725 \AA}$ ($\mathrm{[NII]+H\alpha+[SII]}$) in all spectra. We opt to not mask the $\mathrm{[O II]\,\lambda\lambda3726, 3729}$ doublet in these spectra. While in some non-quiescent cases there is prominent [OII] emission, the total [OII] equivalent widths are typically smaller than those of $\mathrm {[OIII]+H\beta}$ or $\mathrm{[NII]+H\alpha+[SII]}$. Furthermore, due to the coarse wavelength resolution of PRISM around rest $\mathrm{3700 \AA}$, masking this doublet would have reduced spectral sampling of the Balmer break, which is crucial to the selection of quiescent galaxies.

\begin{figure*}[!htb]
    \centering
    \includegraphics[width = 1.\textwidth]{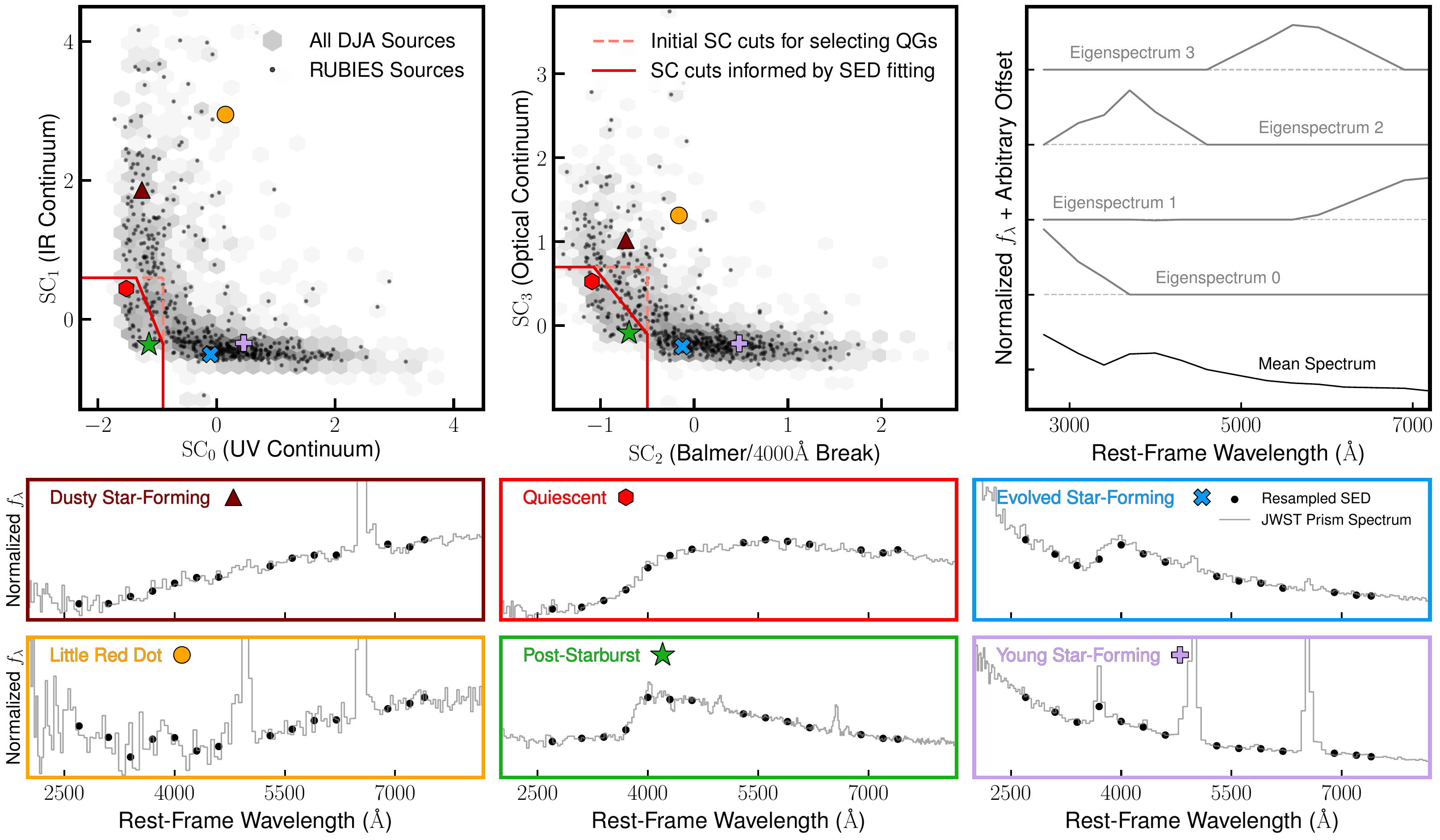}
    \caption{Demonstration of the PCA analysis of NIRSpec/PRISM spectra from the DJA: The top row includes super color (SC) distributions (left and center) and derived eigenspectra (right panel). Each SC corresponds to the normalization of an eigenspectrum for each individual source, thus SC space location maps to spectral types. The bottom rows highlight representative examples: Dusty Star-Forming Galaxies (brown triangle), Little Red Dots (yellow circle), Old Quiescent Galaxies (red hexagon), Young Quiescent/Post-Starburst Galaxies (green star), Evolved/Napping Star-Forming Galaxies (blue cross), and Young Star-Forming Galaxies (purple cross). The de-redshifted original spectrum is shown in grey. The \texttt{SpectRes}-resampled spectrum used in PCA is shown as black dots.  }
    \label{fig:pca_selection}
\end{figure*}

We apply the SparsePCA method implemented by the standard Python package \texttt{Scikit-learn} to our preprocessed spectra. To select the number of eigenspectra $n$ that is the most suitable for our goal of representing and classifying SEDs, we attempt PCA on the same dataset with various $n$ setups. To quantify the quality of the spectroscopic reconstruction, we then define the mean absolute fractional residual of each spectrum as $\langle |(F_{\lambda, original}(\lambda) - F_{\lambda, reconstructed}(\lambda))/F_{\lambda, original}(\lambda) |\rangle$, where the bracket means the average over all wavelength bins. We examine the distribution of the mean fractional residual as a function of $n$. At $n=4$, $99.9\%$ of the sources have a mean fractional residual less than 0.02, suggesting that the reconstructed spectra of these sources account for $\sim 98\%$ of the flux in the original spectra. At $n>4$, each reconstructed spectrum accounts for a higher fraction of flux in the original spectrum, although this improvement becomes marginally better as $n$ grows. At the same time, a higher number of SCs complicates the interpretation of these sources. Therefore, we choose $n=4$ for our subsequent analysis.

The choice of SparsePCA is intended to identify eigen components corresponding to orthogonal features of the spectral continuum. In total, we identify four eigenspectra that correspond to the UV continuum slope, Balmer/$\mathrm{4000 \AA}$ break strength, red optical continuum slope, and IR continuum slope (see Figure \ref{fig:pca_selection} for details). This choice allows us to intuitively connect the distribution of the galaxy population in SC space to their physical properties. Thus, we can draw empirical cuts in the SC space to separate galaxies of different spectral types. 

In Figure \ref{fig:pca_selection}, we showcase our ability to select objects of different spectral types based on their location in the SC space. The vast majority of the galaxies targeted by early public JWST/NIRSpec programs, and therefore in our PCA analysis, are star-forming galaxies (blue and purple). These sources have positive SC$_0$, indicating that they possess excessive rest-frame UV flux on top of the mean spectrum. A fraction of them have a notable Balmer/$\mathrm{4000 \AA}$ break (blue), as suggested by their slightly negative SC$_2$. These galaxies have relatively evolved stellar populations and likely had stochastic SFHs (\citealp{Strait.etal.2023,Looser.etal.2024,Covelo-Paz.etal.2025}; G. Khullar et al. in preparation). The LRDs (yellow), whose physical nature is hotly debated, reside in an isolated regime in the SC space. Since they are typically characterized by their unique ``V-shaped" continua that turn over at the Balmer limit \citep[e.g.,][]{Setton.etal.2024b,Hviding.etal.2025}, these sources tend to have both positive SC$_0$ and SC$_1$. However, the UV colors of faint LRDs are not uniform and consequently, LRDs can shift towards lower SC$_0$ \citep{Williams.etal.2024}. The quiescent galaxies (green and red) have the strongest Balmer/$\mathrm{4000 \AA}$ break strengths and almost no rest-frame UV emission, residing in the lower-left corner of both SC$_0$-SC$_1$ and SC$_2$-SC$_3$ SC space. The spectral signature of older quiescent galaxies (red) is dominated by stars with longer lifetime than A-type stars, as they have typically quenched over $\mathrm{\sim 0.8\,Gyr}$. In contrast, the spectral signature of young quiescent galaxies (or recently quenched/post-starburst galaxies; green) is dominated by A-type stars, since they typically quenched within $\mathrm{\sim 0.8\,Gyr}$. Older quiescent galaxies tend to have flatter optical-near-IR continuum slopes and therefore have slightly higher SC$_3$ and SC$_1$ than the young quiescent galaxies. They are also characterized by a $\mathrm{4000 \AA}$ break instead of the Balmer break characteristic of younger quiescent galaxies, which is reflected by their lower SC$_2$. However, the exact boundary between dusty star-forming galaxies (maroon) and old quiescent galaxies (red) is difficult to determine in the SC space. As the PCA analysis is essentially an operation that lowers the dimension of the dataset, the subtle difference in the spectral shape required to break the dust-age degeneracy becomes hard to map to the SC location. 

Using the visualization of spectral types in Figure \ref{fig:pca_selection}, we can immediately eliminate sources that are unlikely to be quiescent galaxies without having to perform computationally expensive stellar population synthesis modeling on the entire RUBIES dataset. In order to ensure a complete selection of quiescent galaxies in RUBIES, we adopt generous initial SC cuts, outlined by light red dashed lines in Figure \ref{fig:pca_selection}. These cuts effectively remove any sources with spectral features that are mutually exclusive with quiescence. The cut in SC$_0$ removes sources with significant UV flux, which still host massive stars and therefore recent star formation. The cut in SC$_2$ removes sources without a prominent Balmer/$\mathrm{4000 \AA}$ break, which lack evolved stellar populations. The cuts in SC$_1$ or SC$_3$ remove sources with continuum slope at red optical or NIR wavelengths corresponding to significant dust reddening, which is rare in quiescent galaxies \citep{Setton.etal.2024a,Siegel.etal.2025}.  

Using this preselection, we identify an initial sample of 41 unique $\mathrm{m_{F444W} <24}$ sources at $2<z<5$ from RUBIES. Next, we perform spectrophotometric modeling, as follows, to completely disentangle the dusty star-forming and quiescent populations in the initial sample. In Appendix \ref{appendix: effective limit}, we refine these SC cuts to boost selection purity using the best-fitting SED models. The refined SC cuts are shown as red solid lines in Figure \ref{fig:pca_selection}.

\subsection{Spectro-Photometric Fitting}\label{sec:SED fitting}

\begin{figure*}[!htb]
    \centering
    \includegraphics[width = 1.\textwidth]{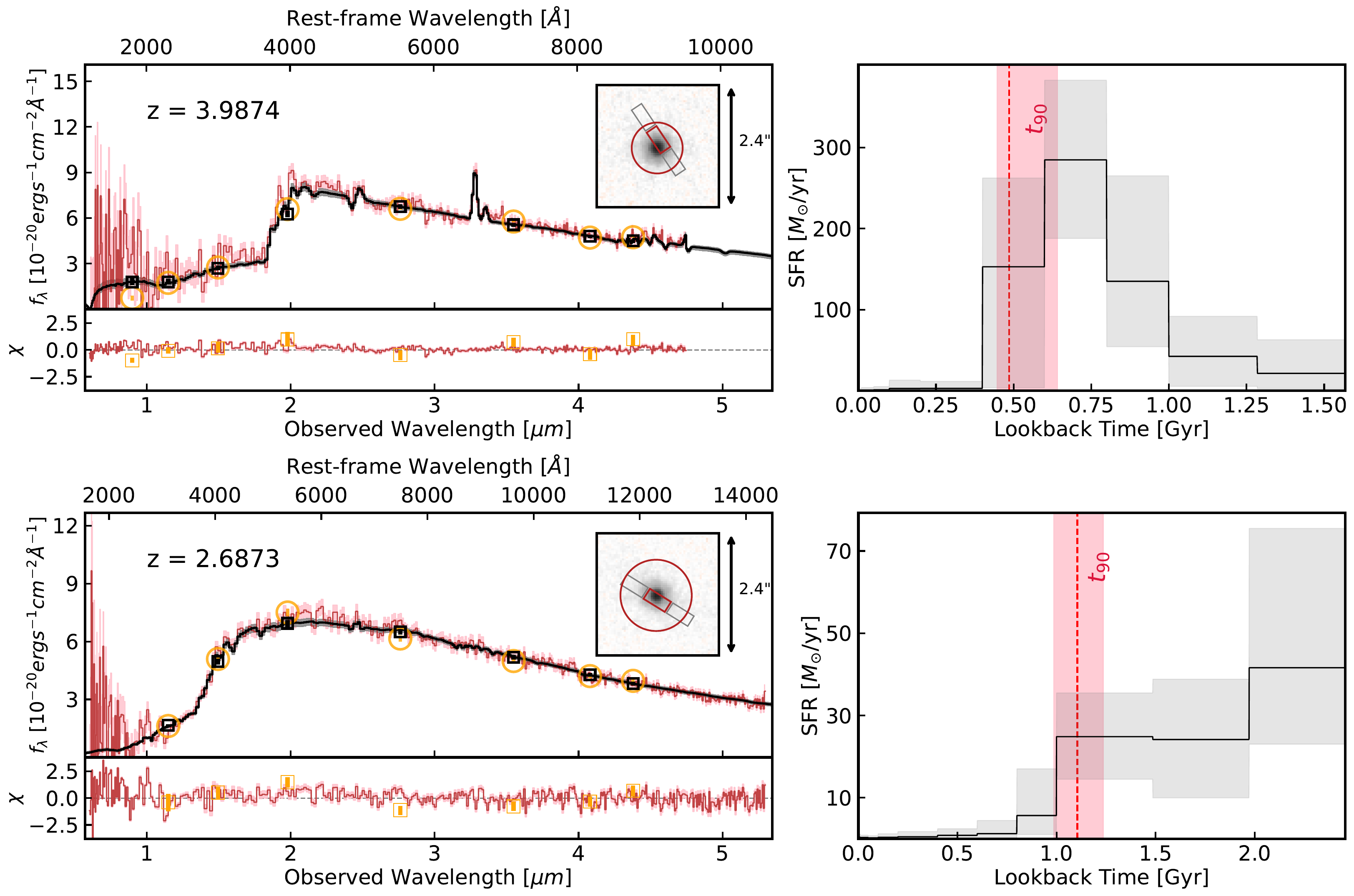}
    \caption{Example spectro-photometric fits of a $z\sim4$ post-starburst galaxy (ID: RUBIES-UDS-12594) and an old quiescent galaxy (ID: RUBIES-EGS-42328). For each row, the upper left panel shows the SED of observed NIRCam photometry and uncertainty (orange circle and error bar), observed NIRSpec PRISM spectrum and uncertainty (red solid line and pink bands), best-fit model photometry and its 68\% confidence interval (black square and errorbar), and best-fit model spectrum and its 68\% confidence interval (black solid line and grey band). The bottom left panel shows the residuals of our fits to the observed photometry (orange squares) and spectrum (red solid line). The inset shows the NIRCam/F444W image of the target galaxy. The red rectangle traces the central MSA micro-shutter used to compute slit-like aperture photometry, and the red circle traces the circular photometric catalog aperture. The right panels show the median (lines) and $16-84\%$ intervals (bands) for the inferred SFHs and $t_{90}$ measurements. These fits allow us to robustly determine the physical properties of these galaxies.}
    \label{fig:SED_young_QG}
\end{figure*}

In order to infer the physical properties of PCA-selected galaxies and finalize our selection of quiescent galaxies, we simultaneously fit the PRISM spectrum and NIRCam photometry, using the Bayesian stellar population inference code \texttt{Prospector} \citep{Leja.etal.2017,Johnson.Leja.2017,Johnson.2021} with the nested sampling code \texttt{dynesty} \citep{Speagle.etal.2020}. \texttt{Prospector} uses the stellar population synthesis models from the Flexible Stellar Population Synthesis (FSPS) package \citep{Conroy.etal.2009,Conroy.Gunn.2010}. We adopt the MILES spectral
library \citep{Sanchez-Blazquez.etal.2006,Falcon-Barroso.etal.2011} and MIST isochrones \citep{Choi.etal.2016,Dotter2016}, assuming a Chabrier IMF \citep{Chabrier.etal.2003}. Using the \texttt{Prospector} \texttt{PolySpecFit} instance, we opt to use a polynomial of order 5 to flux calibrate the spectrum to photometry, which we extract from customized apertures described in Section \ref{Sec: photometry}. We enforce a minimum uncertainty floor of 5\% on both the spectrum and photometry.

The setup strategy in our fits is similar to those in \cite{deGraaff.etal.2025}. We choose a non-parametric SFH that utilizes the continuity prior of Prospector described in \cite{Leja.etal.2019}. Given the wide range of redshifts in our sample, we adopt different age bins for each object. For the most recent $\mathrm{200\,Myr}$ in lookback time, we divide it into 4 bins of $\mathrm{10,40,50,100\,Myr}$, then we linearly add 4 bins of $\mathrm{200\,Myr}$ until we reach $\mathrm{ 1\,Gyr}$ in lookback time. The remaining time is evenly divided into $N_\mathrm{old}$ bins. We calculate $N_\mathrm{old}$ by taking the ceiling of $(t_\mathrm{universe}-1\,\mathrm{Gyr})/0.5\,\mathrm{Gyr}$, where $t_\mathrm{universe}$ is the age of the universe. The number of old bins ranges from 1 to 5 for our sample at $2<z<5$. We fix the redshift to \texttt{msaexp}-derived spectroscopic redshifts. We assume a two-parameter \cite{Kriek.Conroy.2013} dust law, which is parameterized by the attenuation around old ($\mathrm{t > 10 \,Myr}$) stars fit in the range $\tau \in [0,2.5]$ and a free dust index $\delta \in [-1,0.4]$ that describes the deviation from the \cite{Calzetti.etal.2000} dust law and includes a UV bump that depends on the slope parameterized as in \cite{Noll.etal.2009}. We fix the attenuation around young ($\mathrm{t < 10 \,Myr}$) stars to be twice that of the older populations. The stellar metallicity is set as a free parameter with a logarithmically sampled uniform prior in the range $\mathrm{log(Z/Z_{\odot}) \in \,[-2, 0.2]}$. We mask all wavelengths shorter than rest-frame $\mathrm{1200 \AA}$ to avoid contributions from intergalactic medium absorption. We marginalize over the emission lines (specifically, $\mathrm{[Ne V]\,\lambda3426}$, $\mathrm{[O II]\,\lambda\lambda3726, 3729}$, $\mathrm{[O III]\,\lambda\lambda4959,5007}$, $\mathrm{[N II]\,\lambda\lambda6548,6584}$, $\mathrm{[S II]\,\lambda\lambda6716,6731}$, $\mathrm{[S III]\,\lambda\lambda9069,9532}$, $\mathrm{H\alpha}$, $\mathrm{H\beta}$, $\mathrm{H\gamma}$, and $\mathrm{H\delta}$). by fitting Gaussian profiles. In these Gaussian profiles, the normalization is set as a free parameter, the center is fixed to the line center, and the width is tied to the galaxy intrinsic dispersion and convolved with the instrumental resolution.

The line spread function curves in the original JWST User Documentation (JDox) are broader than those measured in practice, depending on the exact source morphology \citep{deGraaff.etal.2024}. Before fitting, all models are convolved with a line spread function that is a factor of 1.3 narrower than the JDox curves to account for instrumental dispersion, following \cite{deGraaff.etal.2025}. We additionally include two free velocity dispersions that smooth the stellar continuum and ionized gas emission, which we allow to vary in the range $\mathrm{[0,1000] \,km/s}$ to marginalize over the uncertainty in the line spread function in addition to the intrinsic dispersion of the galaxy.

As described in Section \ref{Sec: photometry}, physical properties inferred from SED fitting, such as stellar mass, have to be corrected because the customized photometry does not capture the total light from each galaxy due to the small aperture sizes. We derive an aperture to total flux correction from the ratio of F444W flux within the rectangular MSA apertures to the PSF-matched total in the same band. We apply this correction factor to the derived stellar mass and SFH for each galaxy. Once multiplied with the 5th order polynomial and the scaling factor to F444W total aperture flux, the spectrum in these fits typically increases by a factor of $\sim 1-2$, which is a modest correction since the default slit-loss correction in the DJA pipeline cannot be perfect. Following typical conventions, all masses are reported as surviving stellar mass. We note that this modeling assumes solar-scaled stellar population models. Given that these massive galaxies are probably alpha-enhanced \citep[e.g.,]{Beverage.etal.2025} and this will slightly overestimate mass and age \citep{Park.etal.2024}.

In Figure \ref{fig:SED_young_QG}, we show two examples of our spectro-photometric fits: a young quiescent (post-starburst) galaxy at $z\sim4$ (top) and an old quiescent galaxy at $z\sim 2.7$ (bottom). In the upper left panels, the observed spectrum and its uncertainties are shown in red, and the observed photometry with error is shown in orange. The best-fitting model spectrum and photometry are shown in black. Both observed and model spectra have been scaled by the polynomial calibration vector to match the photometry. The insets show the NIRCam/F444W cutout ($2.4''$ in width) of the target galaxy. The red rectangle traces the central MSA micro-shutter (for slit-like aperture photometry), and the red circle traces the circular aperture from which the catalog photometry was extracted. The bottom left panels show the residuals for the spectrum and photometry of the fits ($\chi$ values), which we define as the difference between observed flux and best-fitting model flux, normalized by the uncertainties in observed flux. The right panels show the median and $68\%$ confidence interval SFHs for these galaxies as solid black curves and grey regions. We show the median and $68\%$ confidence interval of $t_{90}$ (lookback time at which 90\% of the current stellar mass was formed) in the same panels as dashed red lines and light red regions. Overall, these fits are excellent and enable the characterization of SFHs and the physical properties of these galaxies. The SED fits and SFHs of the remaining galaxies in this sample are shown in the Appendix.

\section{The Selection of Quiescent Galaxies} \label{Sec: Results}
\subsection{The Selection and Classification of the RUBIES Quiescent Galaxy Sample} \label{sample definition}

 \begin{figure*}[!htb]
    \centering
    \includegraphics[width = 1.\textwidth]{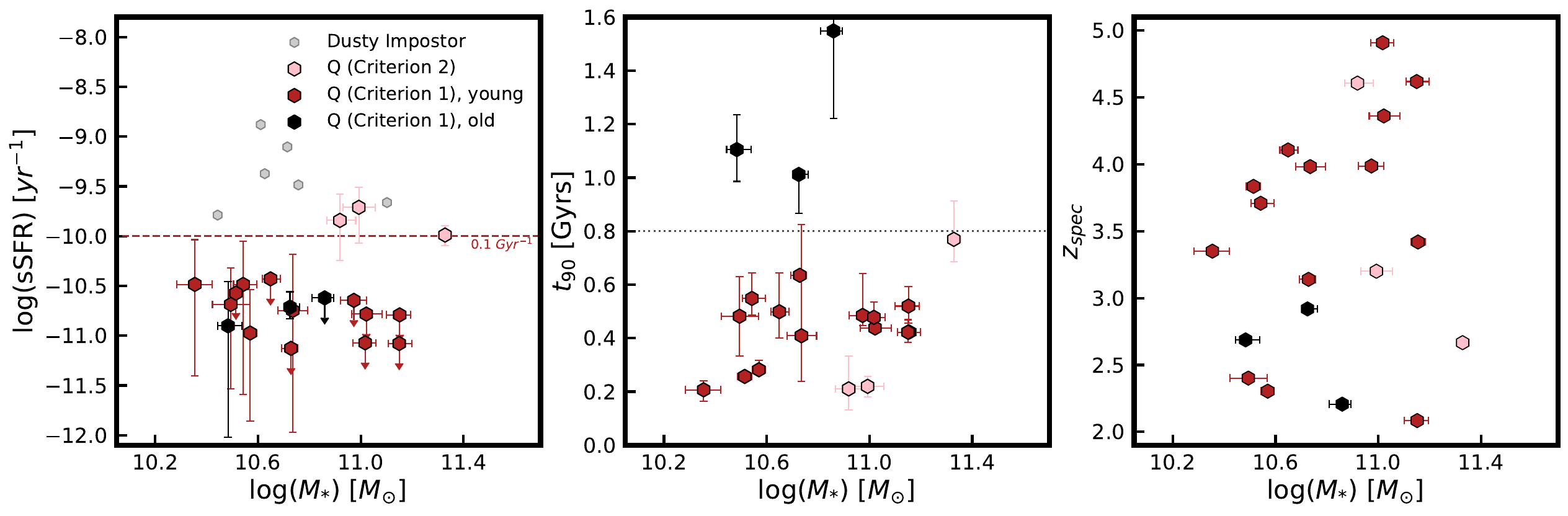}
    \caption{Left panel: Stellar mass versus sSFR for SC-selected quiescent galaxy candidates. Robust quiescent galaxies (Criterion 1) are shown in red, and marginally quiescent cases (Criterion 2) are shown in pink. Galaxies with the 50th percentile $\mathrm{ sSFR} < 10^{-10}\, \mathrm{yr}^{-1}$ are shown as upper limits. One galaxy (ID: RUBIES-EGS-61168) is not in the range of this plot due to its extremely low sSFR. Middle panel: stellar mass versus $\mathrm t_{90}$ for all galaxies. We additionally divide our sample into young quiescent galaxies ($\mathrm t_{90} < 0.8$\, Gyr, red or pink) and old quiescent galaxies ($\mathrm t_{90} > 0.8$\, Gyr, black). Right panel: Stellar mass versus spectroscopic redshift of the finalized sample galaxies. The old quiescent population emerges at $z \sim 3$.}
    \label{fig:SED_selection}
\end{figure*}

Based on our initial PCA selection, we find that that the majority ($\sim 90\%$) of the bright ($\mathrm{F444W<24}\, \cup\,\mathrm{SNR >5}$) RUBIES sources at $2<z<5$ possess empirical spectral features that unambiguously eliminate their possibility of being quiescent systems (see Section \ref{sec:pca_selection} for a qualitative discussion). Our approach to identifying quiescent galaxies is iterative. First, we apply the initial conservative SC cuts (light red dashed lines in Figure \ref{fig:pca_selection}) to focus on potentially quiescent sources and minimize the computational cost. We model the pre-selected galaxies with \texttt{Prospector} and infer their physical properties. We examine the relation between the SCs and $\mathrm{sSFR}$ among these galaxies and identify a smaller region in SC space that robustly identifies quiescent galaxy SEDs (red solid lines in Figure \ref{fig:pca_selection}). This selection largely avoids contamination from other red sources, including dusty star-forming galaxies and those with decreased, but significant, ongoing star formation\footnote{This refined SC selection misses only one marginally quiescent (defined later in text) galaxy (ID: RUBIES-EGS-37883), whose SED is difficult to model with \texttt{Prospector} in the current setup. This galaxy has excessive rest-frame NIR fluxes compared to the best-fitting model, which is likely caused by AGN-induced dust re-emission.}. Next, we describe the identification and classification of quiescent galaxies, using the physical properties and SFHs of galaxies selected by the refined SC cuts. 

In the left panel of Figure \ref{fig:SED_selection}, we show the best-fitting stellar mass versus sSFR from \texttt{Prospector} for all galaxies in the refined SC-selected sample. To test whether binary classification is sensitive to measurement uncertainties, we define two quiescent criteria as follows:
\begin{itemize}
  \item{Criterion 1 (red symbols): $\mathrm{sSFR_{50th} < 10^{-10}\, yr^{-1}}$}
  \item{Criterion 2 (pink symbols): $\mathrm{sSFR_{16th} < 10^{-10}\, yr^{-1}}$}
\end{itemize}
We adopt Criterion 1 as our primary criterion of quiescence, which yields 17 quiescent galaxies in total as our fiducial sample. Criterion 2 includes three more galaxies that could have been selected by Criterion 1 if their $\mathrm{sSFR}$ are perturbed by one $\sigma$. In the following discussion, we also refer to these three galaxies as ``marginally quiescent". We verify that an evolving sSFR threshold (e.g., $\mathrm{sSFR_{50th} < 0.2/t_{universe}(z)\, Gyr^{-1}}$ as in \citealp{Carnall.etal.2023,Baker.etal.2024}) identifies almost the same quiescent galaxies as our fiducial sample, including one additional galaxy at $4<z<5$ (ID: RUBIES-UDS-140707). 

Using the SFHs inferred by \texttt{Prospector}, we calculate the formation time scale ($\mathrm{t_{90}}$ or the lookback time by which a galaxy formed $90\%$ of its current stellar mass) of each quiescent galaxy. In the following analysis, we use $\mathrm{t_{90}}$ as a proxy for time since quenching to differentiate between the old and young galaxy populations. In the middle panel of Figure \ref{fig:SED_selection}, we show $\mathrm{t_{90}}$ versus stellar mass for all quiescent galaxies. Most of the quiescent galaxies in our sample are young (pink or red) with best fit $\mathrm{t_{90}< 0.8\, Gyr}$. In the right panel of Figure \ref{fig:SED_selection}, we show the distribution in stellar mass versus redshift. All older ($\mathrm{t_{90}> 0.8\, Gyr}$) sources (black symbols) are found at the lowest redshifts ($2<z<3$). The SEDs of these galaxies prominently display $\mathrm{ 4000 \AA}$ breaks rather than Balmer breaks (see Figure \ref{fig:SED_young_QG} and Figure \ref{fig:all_gallery01}). The earliest emergence of the older quiescent population at $z\sim3$ in this sample is roughly contemporaneous with the record holder of the ``maximally old" quiescent galaxy at $z = 3.2$ reported in \cite{Glazebrook.etal.2024}. At face value, they are likely the descendants of the first generation of massive quiescent galaxies at $z \geq 5-7$ \citep{Weibel.etal.2024,deGraaff.etal.2025}. However, we note that the stellar masses of these examples are slightly lower than the current higher-redshift counterparts in \cite{Weibel.etal.2024} and \cite{deGraaff.etal.2025}. This discrepancy could simply reflect cosmic variance or a systematic bias in mass inference between galaxies in different evolution stages. The inferred physical properties relevant to the selection and classification of the complete quiescent sample are tabulated in Table \ref{tab:all_measurements}. 

This final sample includes the most massive quiescent galaxies (median $\mathrm{M_{*} \sim 10^{10.7}M_{\odot}}$) at $2<z<5$ in RUBIES, in part because we enforce the parent sample to include only the bright galaxies with $\mathrm{m_{F444W} < 24}$. However, this magnitude limit does not translate to a uniform mass limit for the entire final sample. Although these galaxies span a wide range of redshifts ($2<z<5$), galaxies are younger and therefore have higher mass-to-light ratios at higher-redshift, somewhat mitigating the increasing luminosity distance. 

To derive the redshift-dependent effective mass-completeness limit, we obtain rescaled analogs of the observed galaxies, using a grid of redshifts and stellar masses and best-fitting \texttt{Prospector} models. For a group of models within a range of redshift, we compute their $\mathrm{F444W}$ magnitude corresponding to a given stellar mass and examine where they would pass the F444W$<24$ threshold. The details of this analysis are presented in Appendix \ref{appendix: effective limit}. It suggests that our magnitude-limited selection is complete above $\mathrm{10^{10.3}M_{\odot}}$ at $2<z<3$, above $\mathrm{10^{10.5}M_{\odot}}$ at $3<z<4$, and above $\mathrm{10^{10.6}M_{\odot}}$ at $4<z<5$.

We also investigate how dust attenuation ($A_V$) could affect the locations of the true quiescent population in this sample in SC space. Using the best-fit \texttt{Prospector} models for all $\mathrm{sSFR_{50th}} < 10^{-10}\, \rm yr^{-1}$ galaxies, we re-generate model SEDs for a grid of $A_V$ values. The details of this analysis are presented in Appendix \ref{appendix: effective limit}. In brief, the true quiescent galaxies generally move along a diagonal track with a negative slope in SC$_0$-SC$_1$ and SC$_2$-SC$_3$ as their dust attenuation increases. In addition to the initial SC thresholds, the diagonal SC cuts in SC$_0$-SC$_1$ and SC$_2$-SC$_3$ (shown as red solid lines in Figure \ref{fig:pca_selection}) are roughly parallel to the $A_V$ evolution track of the true quiescent population. We find that our refined SC cuts should be complete for quiescent galaxies with $A_V<0.7$.

\subsection{How Would This Sample Have Been Selected with Rest-frame Colors?}\label{Sec: probe UVJ}
\begin{figure*}[!htb]
    \centering
    \includegraphics[width = 1\textwidth]{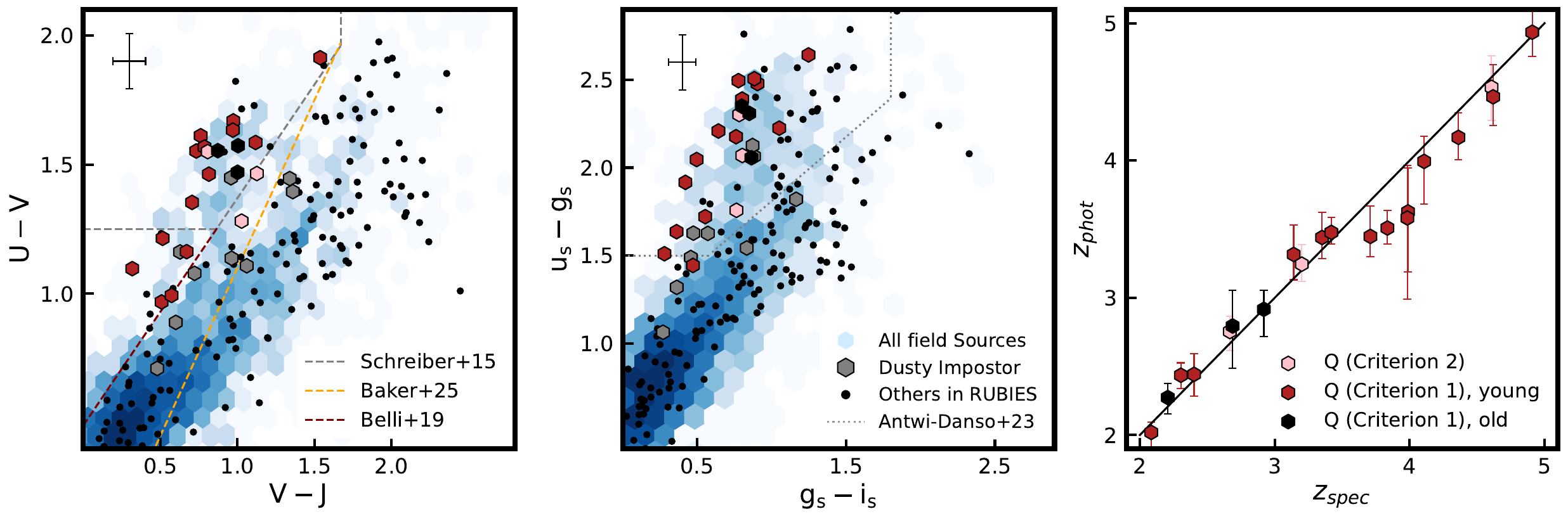}
    \caption{Left panel: Rest-frame {\it UVJ} colors of all ($\mathrm{F444W<24}$) galaxies at $2<z<5$ in EGS and UDS. We show the SC-selected RUBIES galaxies as hexagon symbols, whose color coding follows previous convention. We show the remaining galaxies in RUBIES as black dots. Finally, the distribution of any EGS or UDS galaxies not in RUBIES are shown by 2D histogram bins, in which the darker color indicates higher density. Middle panel: Rest-frame {\it $u_s$$g_s$$i_s$} colors of the same set of galaxies. The characteristic uncertainties in these rest-frame colors are shown by the error bars in the upper left corners. All rest-frame color selections shown here suffer from various degrees of impurity and incompleteness, which we discuss in detail in the text. Right panel: EAZY-derived photometric redshifts versus spectroscopic redshifts of all RUBIES massive quiescent galaxies. These photometric redshifts are largely consistent with the redshifts recovered from PRISM spectroscopy, though they are the main contribution to uncertainties in rest-frame colors. }
    \label{fig:demographic}
\end{figure*}

While spectroscopy remains the most robust tool to identify quiescent galaxies, photometric samples will always provide a comprehensive and cost-efficient probe of the cosmic volume, although at the expense of purity \citep[e.g.,][]{Carnall.etal.2023,Valentino.etal.2023,Long.etal.2024,Alberts.etal.2024}. We now estimate the purity and completeness of various rest-frame color selection criteria if applied to this spectroscopic sample of massive quiescent galaxies at $2<z<5$. 
 
In Figure \ref{fig:demographic}, we show the rest-frame {\it UVJ} color distribution of all galaxies in EGS and UDS with $\mathrm{F444W<24}$ at $2<z<5$ in the left panel and their rest-frame {\it $u_s$$g_s$$i_s$} colors \citep{Antwi-Danso.etal.2023} in the middle panel. The galaxies in RUBIES that are selected by our refined SC cuts are shown as hexagon symbols, whose color coding follows previous convention. Other galaxies in RUBIES that are excluded by the refined SC cuts are shown as black dots. All the remaining field galaxies not surveyed by RUBIES are binned into 2D histograms, where the darker color indicates the higher density of objects.

The conventional rest-frame {\it UVJ} selection criteria with a horizontal {\it U-V} cut \citep[e.g.,][]{Williams.etal.2009,Whitaker.etal.2011,Schreiber.etal.2015} misses some of the youngest quiescent galaxies in our sample, as expected \citep[e.g.,][]{Baker.etal.2024}. Extending to bluer {\it U-V} colors \citep{Belli.etal.2019,Baker.etal.2024} (diagonal dashed lines) captures many excluded galaxies, for this quiescent sample ($\mathrm{sSFR_{50th}} < 10^{-10} \,yr^{-1}$) $\sim 30\%$ have ${\it U-V} < 1.23$. We find contamination fractions of $\sim 35\%$ in both the classical and extended {\it UVJ} criteria, which is high but largely consistent with the $\sim 10\%-30\%$ reported in \cite{Leja.etal.2019b}, \cite{Antwi-Danso.etal.2023}, or \cite{Nanayakkara.etal.2025}. We return to discuss the impact of false positivity on number densities from {\it UVJ} color selections in Section \ref{fig:number-densities-observations}. Alternatively, the {\it $u_s$$g_s$$i_s$} criteria select almost all ($\sim 95\%$) quiescent galaxies in this sample at the cost of much higher contamination rates ($\sim 60\%$). 

Even though the photometric redshifts of these bright quiescent galaxies agree well with the spectroscopic redshifts (see right panel of Figure \ref{fig:demographic}), the substantial uncertainties in photometric redshifts can contribute significant systematic uncertainties to the rest-frame color estimations and scatter in color-color space. The typical uncertainties in these color estimations are shown in the left and middle panels of Figure \ref{fig:demographic}. We emphasize that this scatter is not included in our purity and completeness fractions quoted above. Both the impostors and true quiescent galaxies located near these selection boundaries can shift into or out of these color selection criteria if perturbed within the confidence interval of their rest-frame colors. These uncertainties and the nuance in the definition of true quiescence (Criterion 1 versus Criterion 2) complicate the determination of systematic impurity or incompleteness in these rest-frame color selections. 

Another caveat is that the completeness or purity fraction reported in this section only considers the EGS and UDS galaxies surveyed by RUBIES and is not corrected for the selection bias introduced by the RUBIES selection function. Without this correction, the purity fraction of the extended {\it UVJ} criteria is likely prone to underestimation. These criteria include younger quiescent galaxies but also bluer impostors \citep{Whitaker.etal.2012,Spitler.etal.2014,Straatman.etal.2015}. Since RUBIES preferentially targeted redder sources, which we discuss in detail in the following section, RUBIES could miss more blue impostors and underestimate the total number of sources present in the extended color region.

\subsection{Correcting for Targeting Incompleteness with the RUBIES Selection Function} \label{Sec: Correcting for Incompleteness}

\begin{figure*}[!htb]
    \centering
    \includegraphics[width = 1.\textwidth]{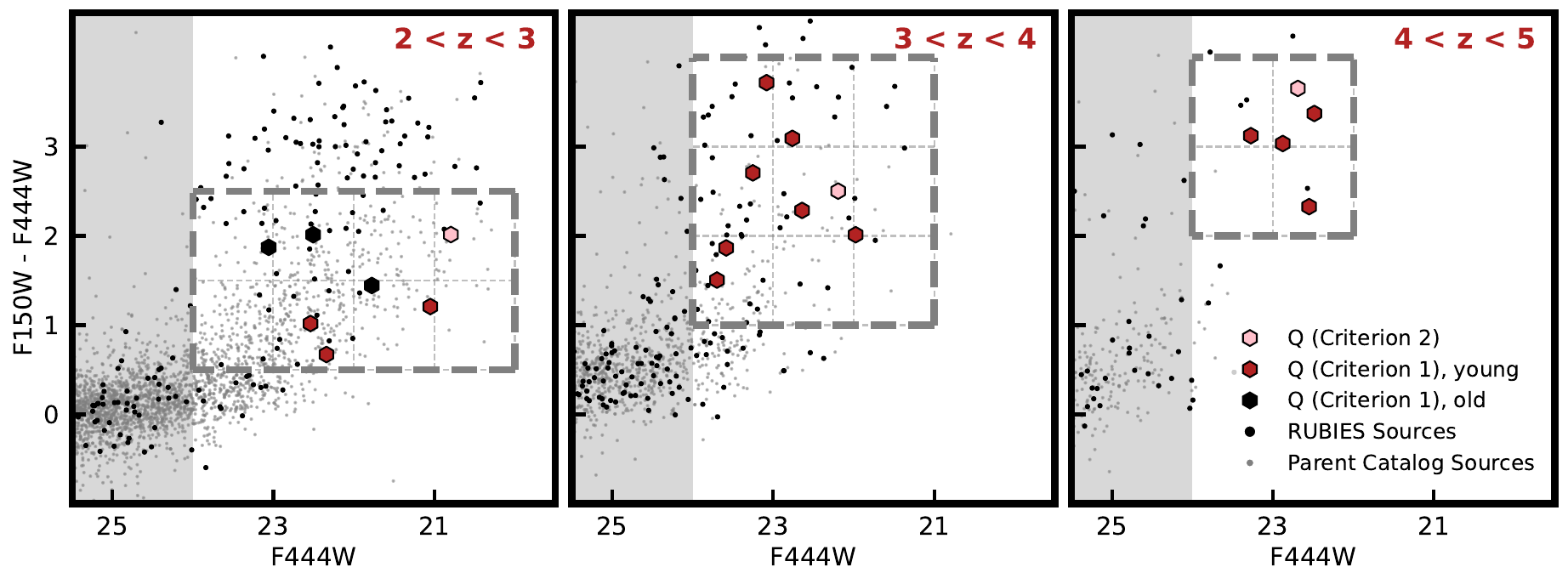}
    \caption{The parent catalog F444W magnitude versus F150W-F444W color of this quiescent sample, RUBIES surveyed sources, and all parent catalog sources in PRIMER/UDS and CEERS/EGS in three redshift bins. These panels illustrate the survey completeness of RUBIES as a function of magnitude, color, and redshift. The boxes outlined by grey dashed lines represent the area in color-magnitude space in which we perform the completeness calculation in each redshift bin. The survey completeness is close to $100\%$ for sources under the footprint in the color-magnitude space occupied by massive quiescent galaxies at $z>4$.}
    \label{fig:color-mag}
\end{figure*}
Spectroscopy reliably prevents the contamination of dusty star-forming galaxies in the selection of quiescent galaxies, which is unavoidable with only photometry. However, these spectroscopic surveys cannot comprehensively target every eligible source in the field. The number densities derived from this kind of spectroscopic selection must be statistically corrected for sample incompleteness. For the RUBIES massive quiescent galaxy sample, the overall sample completeness is a net product of three selection steps.

Firstly, we consider the fraction of field sources that are eligible to be included as survey candidates ($\mathrm{S_{el}}$). This is a function of the apparent photometric properties of the source $\vec{m} \mathrm{= ( m_{F444W}, m_{F150W} - m_{F444W})}$, which depends on their intrinsic physical properties such as stellar mass ($\mathrm{M_{*}}$), sSFR, redshift ($\mathrm{z}$), and dust reddening ($\mathrm{A_V}$). Secondly, we define a variable, $\mathrm{S_{slit}}$, to be the fraction of potential sources for which suitable spectra were obtained. This fraction includes whether a source is assigned an MSA slit, whether a spectrum was successfully obtained and reduced, and whether the spectroscopic wavelength coverage, redshift quality grade, and SNR meet the PCA requirements described in Section \ref{sec:pca_selection}. This is similarly a function of the photometric properties of the source $\vec{m}$. Thirdly, we consider the fraction of sources included in our super color cuts among all potential quiescent sources for which a successful spectrum is obtained($\mathrm{S_{PCA}}$). This term is a function of the source spectrum $\mathrm{f_{\lambda}}$ and accounts for missing potential quiescent galaxies with unusual SCs. In summary, we express these terms as:
\begin{itemize}
\item{$\mathrm{S_{el}}(\vec{m}\mathrm{(M_{*}, sSFR,z, A_V)}) $}
\item{$\mathrm{S_{slit}}(\vec{m}\mathrm{(M_{*}, sSFR,z, A_V)}) $}
\item{$\mathrm{S_{PCA}}(f_{\lambda}\mathrm{(M_{*}, sSFR,z, A_V)}) $}
\end{itemize}
Therefore, the total sample completeness is given by:
\begin{equation}
 \mathrm{S_{tot} (M_{*}, sSFR,z, A_V)= S_{el}\cdot S_{slit}\cdot S_{PCA}}.
\end{equation}

We simplify some of the terms in the selection function as follows. Since RUBIES considers any field source with $\mathrm{F444W<28.5}$ to be eligible and we have enforced all sources in this sample to have $\mathrm{m_{F444W}<24}$, we assume $\mathrm{S_{el}} = 1$ for all galaxies in this sample. Furthermore, $\mathrm{S_{PCA}}$ has no dependency on stellar mass or redshift, since our PCA analysis is conducted on normalized, rest-frame spectra. As we have shown, the SC selection of quiescent galaxies has an effective limit in dust attenuation of $\mathrm A_V <0.7$, we assume $\mathrm{S_{PCA}}\sim 1$ at $\mathrm{A_V} <0.7$ and $\mathrm{sSFR < 10^{-10} \, yr^{-1}}$. With these simplifications, we show that $\mathrm{S_{tot}}$ only depends on $\mathrm{S_{slit}}$ for the quiescent galaxies in our selection.

RUBIES selects spectroscopic targets from eligible sources in CEERS/EGS and PRIMER/UDS according to their photometric redshifts, F444W magnitudes, and $\mathrm{F150W-F444W}$ colors \citep{deGraaff.etal.2024rubies}. To estimate $\mathrm{S_{slit}}$, we divide the finalized quiescent galaxies, RUBIES sources with successful spectra, and eligible parent sources into three redshift bins: $[2,3], [3,4], $ and $[4,5]$. In Figure \ref{fig:color-mag}, we show the color-magnitude distribution of finalized quiescent galaxies (using the same color coding as in previous figures), RUBIES sources with successful spectra (black dots), and the parent photometric catalog sources (grey dots) in each redshift bin. We select the approximate color-magnitude space occupied by the quiescent galaxies in this sample in each redshift bin (thick grey dashed line) and divide these color-magnitude spaces into smaller boxes that are $\mathrm{1 \,mag}$ in width (thin grey dashed line). We mask out the faint regime ($\mathrm{m_{F444W}>24}$), which is excluded by the quiescent galaxy selection in this sample. In each color-magnitude box ($\mathbf{m}$), we count the total number of quiescent galaxies in this sample $\mathrm{N_{QG, surveyed,\mathbf{m}}}$ (hexagons), the number of RUBIES targets with a successful spectrum ($\mathrm{N_{surveyed, \mathbf{m}}}$; black dots), and the number of eligible parent sources ($\mathrm{N_{total,\mathbf{m}}}$; gray dots). The fraction of eligible sources which yield successful spectra varies smoothly in the color-magnitude space \citep{deGraaff.etal.2024rubies}. This allows us to approximate this fraction as a constant per color-magnitude box: $\mathrm{S_{slit}}(\vec{m}) \mathrm{\sim N_{surveyed,\mathbf{m}}/N_{total,\mathbf{m}}}$, when $\vec{m} \in \mathbf{m}$.

A few sources selected by RUBIES to be assigned an MSA slit fail to produce a spectrum that meets the PCA pre-requisite in wavelength coverage, redshift quality grade, or SNR. However, we find that this failure rate does not have a significant dependence on color, magnitude, or redshift within the relevant parameter space. Generally, the spectroscopic redshifts recovered by RUBIES in the entire survey agree well with photometric redshifts (the normalized median absolute deviation between the two is small: $\sigma(\Delta z/(1 + z)) = 0.033$) \citep{deGraaff.etal.2024rubies}. For the bright quiescent galaxies in this sample, the match between photometric and spectroscopic redshifts is excellent, as shown in the right panel of Figure \ref{fig:demographic}. We assume that any remaining unsurveyed quiescent galaxies should also have reliable photometric redshifts. Therefore, the approximation described above is likely not affected by any catastrophic errors in photometric redshifts or biased by any potential correlation between spectrum acquisition failure and source color, magnitude, or redshift.

For a population of RUBIES massive quiescent galaxies in a narrow redshift bin, which have $ \mathrm{sSFR < sSFR_{max}}$, $\mathrm {M_{*}> M_{*,min}}$, and $\mathrm{A_V < A_{V,max}}$, the number density is given by their expected number divided by the effective survey volume:
\begin{equation}
\begin{split}
 \mathrm{n( M_{*}>M_{*,min}, sSFR< sSFR_{max}, A_V < A_{V,max})} &
 = \\ \mathrm{\frac{1}{V_{eff}} \int^{\infty}_{M_{*,min}}   \int^{sSFR_{max}}_{-\infty}  \int^{A_{V,max}}_{0}} \\ 
 \mathrm{\frac{\Phi(M_{*}, \mathrm{sSFR}, A_V)}{S_{tot} (M_{*}, \mathrm{sSFR}, A_V)} dM_{*} \, d\mathrm{sSFR} \,dA_V},
\end{split}
\end{equation}
where $\mathrm{\Phi(M_{*}, sSFR, A_V)}$ is the number of RUBIES galaxies per stellar mass per sSFR per attenuation per comoving cosmic volume. We do not intend to solve for this integral in the intrinsic parameter space, since the true distribution of $\mathrm{\Phi(M_{*}, sSFR, A_V)}$ is challenging to constrain. We instead assume that the above integral in the intrinsic parameter space (within the effective selection limits) is equivalent to the following summation in the observed parameter space:
\begin{equation} \label{eq: number density}
\mathrm{n_{Massive,Q}  =\frac{1}{V_{eff}} \sum^{\mathbf{m}} \frac{N_{QG, surveyed,\mathbf{m}}\cdot N_{total,\mathbf{m}}}{N_{surveyed,\mathbf{m}}}},
\end{equation}
where $\rm n_{Massive,Q}$ is the number density of massive quiescent galaxies.
And $\mathrm{V_{eff}}$ is given by:
\begin{equation}
 \mathrm{V_{eff} = \frac{\Omega_{field}}{3}\cdot[d_{com}(z_{max})^{3}-d_{com}(z_{min})^{3}]},
\end{equation}
where $\rm \Omega_{field}$ is the angular size of the effective survey area and $\rm d_{com}(z)$ is the co-moving distance to redshift $\rm z$.

This approximation assumes that any source with similar intrinsic properties ($\mathrm{M_{*}, sSFR, A_V}$) should display similar observed source properties ($\mathrm{m_{F444W}, m_{F150W} - m_{F444W}}$), in the special case of massive ($\mathrm{M_{*} >10^{10.3}M_{\odot}}$) quiescent ($\mathrm{sSFR <10^{-10}yr^{-1}}$) galaxies without significant dust reddening ($\mathrm{A_V < 0.7}$). We essentially estimate how the selection of quiescent galaxies in RUBIES depends on their intrinsic source properties, given that we know how the selection of RUBIES sources depends on their observed source properties.

\section{Quiescent Galaxy Number Densities} \label{Sec: Number Density}

\subsection{The Number Density of Massive Quiescent Galaxies in the RUBIES}

\begin{figure*}[!htb]
    \centering
    \includegraphics[width = 1.\textwidth]{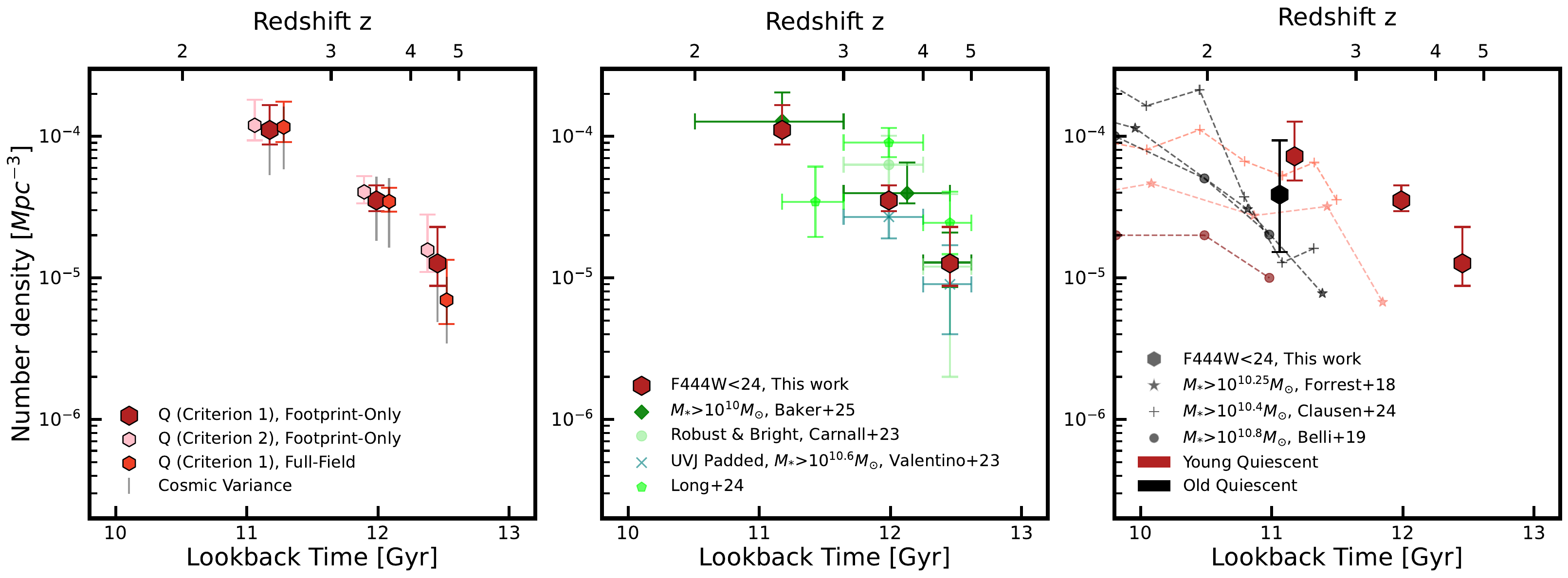}
    \caption{Left panel: Different flavors of total quiescent galaxy number density binned by redshifts from this sample. We show the number density derived from the entire extended quiescent sample (pink hexagons), the entire strict quiescent sample (red hexagons), and the strict quiescent sample but using parent catalog sources from the entire field rather than just the survey footprints for completeness calculation (orange hexagons). Neither our quiescence criteria nor any potential survey targeting bias has a significant impact on the resulting number density. Middle panel: Our fiducial quiescent galaxy number density compared to literature values using JWST data. Our result is in agreement with previous JWST measurements overall. Right panel: Number densities of old ($t_{90} > 0.8\,Gyr$) and young ($t_{90} < 0.8\,Gyr$) quiescent galaxies in this work compared to (non-JWST) literature values. We note that for these comparisons, the number densities derived from this work only adopt a magnitude limit $\rm F444W<24$ and no additional mass limit. This work finds a higher number density for young quiescent galaxies at $3<z<4$ and for old quiescent galaxies at $2<z<3$ than previous non-JWST works. }
    \label{fig:number-densities-observations}
\end{figure*}

We show the resulting comoving number densities in the left panel of Figure \ref{fig:number-densities-observations} in three flavors. The RUBIES survey footprints overlap half of the sky covered by JWST NIRCam imaging in EGS and UDS \citep{deGraaff.etal.2024rubies}. In part to maximize survey efficiency, the MSA mask design of RUBIES targets over-densities within these fields and could result in biases in our number density estimations. To probe how potential cosmic over-densities under the RUBIES footprint can impact our conclusions, we take the fiducial quiescent sample ($\mathrm{sSFR_{50th}} < 10^{-10} \, \mathrm{yr^{-1}}$) and calculate the number density in each redshift range using two different approaches. In the first approach, we only count parent catalog sources that fall under the survey footprint for completeness estimation. In this case, we divide the resulting expected number of quiescent galaxies by the comoving volume corresponding to only the total sky area under the footprints ($\mathrm{\Omega_{field}} \sim 150 \, \mathrm{arcmin}^2$). In the second approach, we count sources from the entire UDS/EGS footprints as $\mathrm{N_{total}}$ for completeness estimation. In this case, we use the comoving volume corresponding to the entire sky area of the NIRCam imaging ($\mathrm{\Omega_{field} \sim 304 \, arcmin^2}$). As shown in the left panel of Figure \ref{fig:number-densities-observations}, the targeting bias due to over-densities at $4<z<5$ results in a footprint-only number density (red) $\sim 0.2 \,\mathrm{dex}$ higher than that of the full field (orange). This offset effectively reflects the cosmic variance within EGS and UDS.

To test the impact of different quiescent criteria in Section \ref{sample definition}, we calculate a third flavor with the extended quiescent sample ($\mathrm{sSFR_{16th} < 10^{-10} \, yr^{-1}}$), adopting the ``footprint-only" approach. An extended criterion for quiescent galaxy selection (Criterion 2) slightly increases the resulting number density (pink; Figure \ref{fig:number-densities-observations}, left panel) at all redshifts by $\sim 0.1 \,\mathrm{dex}$. We tabulate the various flavors of quiescent galaxy number densities in Table \ref{tab:number densities}. Overall, these variations are insignificant compared to uncertainties due to cosmic variance, which we will describe in the following. We choose to report the number densities calculated with the ``Criterion 1" sample and the ``footprint-only" approach as our fiducial result.

\begin{table*}[thbp]
\caption{RUBIES Massive Quiescent Galaxy Number Densities}
    \centering
    \begin{tabular}{c|ccc}
    \hline
    \hline
       Mass or Magnitude limit& $\mathrm{m_{F444W} <24}$ & $\mathrm{m_{F444W} <24}$ & $\mathrm{M_{*} > 10^{10.5}M_{\odot}}$   \\
       Quiescence Criterion & $\mathrm{sSFR_{50th} < 10^{-10} \, yr^{-1}}$ & $\mathrm{sSFR_{16th} < 10^{-10} \, yr^{-1}}$ & $\mathrm{sSFR_{50th} < 10^{-10} \, yr^{-1}}$   \\
     \hline
      $2<z<3$& $1.10^{+0.56}_{-0.24} \cdot10^{-4}\mathrm{Mpc^{-3}}$& $1.20^{+0.61}_{-0.26} \cdot10^{-4}\mathrm{Mpc^{-3}}$ & $0.77^{+0.48}_{-0.19} \cdot10^{-4}\mathrm{Mpc^{-3}}$ \\
      $3<z<4$&$3.53^{+0.97}_{-0.57} \cdot10^{-5}\mathrm{Mpc^{-3}}$ &$4.05^{+1.17}_{-0.68} \cdot10^{-5}\mathrm{Mpc^{-3}}$&$2.74^{+0.86}_{-0.48} \cdot10^{-5}\mathrm{Mpc^{-3}}$ \\
      $4<z<5$&$1.26^{+1.01}_{-0.38}\cdot10^{-5}\mathrm{Mpc^{-3}} $ & $1.57^{+1.21}_{-0.47} \cdot10^{-5}\mathrm{Mpc^{-3}}$& $1.26^{+1.01}_{-0.38} \cdot10^{-5}\mathrm{Mpc^{-3}}$
      
    \end{tabular}
    \label{tab:number densities}
\end{table*}

To estimate the measurement uncertainties on the derived number densities, we calculate the binomial proportion confidence interval for the spectroscopic completeness in each color-magnitude box ($\mathrm{ N_{surveyed,\mathbf{m}}/N_{total,\mathbf{m}}}$), following the method in \cite{Wilson.1927} (Wilson score interval). We propagate the confidence interval in completeness into number densities, using Equation \ref{eq: number density}. We plot the resulting uncertainties due to counting and completeness corrections as colored error bars in the left panel of Figure \ref{fig:number-densities-observations}. 

Additionally, we estimate the systematic uncertainty due to cosmic variance using a method similar to that described in \cite{Taylor.etal.2022}. Briefly, we use $\sim 1300 -2300$ mock light cones of similar area to the RUBIES survey derived from the UniverseMachine model \citep{Behroozi.etal.2019} run on the MDPL2 dark-matter-only N-body simulations \citep{Reibe.etal.2013}. Mock galaxies are rank-ordered by stellar mass, and comparison samples are abundance-matched (to observed number densities in each redshift range) to estimate the 1-$\sigma$ scatter in the field-to-field variation. Due to the small survey volume of RUBIES, the estimated cosmic variance is large (($\sim 60\%$) and dominates the uncertainty budget for this sample.

\subsection{Comparing to Other Observational Results}

In the middle panel of Figure \ref{fig:number-densities-observations}, we compare our fiducial quiescent galaxy number densities to other JWST observations. Overall, our measurements are in good agreement with other JWST results. Notably, \cite{Baker.etal.2024}, which is the only other spectroscopic sample shown, is the most consistent with our results at $2<z<4$. That work combined multiple NIRSpec observation programs \citep{Eisenstein.etal.2023a,Eisenstein.etal.2023b} targeting the GOODS fields \citep{Giavalisco.etal.2004} and utilizes a similar approach: pre-selecting quiescent candidates with rest-frame UVJ colors (with an extended color cut) and finalizing with results from spectro-photometric modeling. They report five quiescent galaxies at $4<z<4.5$ but none above $z>4.5$. Taken together with our results, this suggests that the number density of quiescent galaxies likely drops steeply at $4<z<5$. 

The other literature samples shown in the middle panel of Figure \ref{fig:number-densities-observations} are derived from photometric samples. \cite{Carnall.etal.2023} selected quiescent galaxies based on sSFR inferred from photometry-only SED fits, using the NIRCam imaging data of CEERS/EGS (which partially overlaps with the RUBIES footprint); we show the ``robust, bright" sub-sample \footnote{RUBIES spectra reveal that one of the ``robust" quiescent galaxies in \cite{Carnall.etal.2023} is still star-forming while one non-``robust" quiescent galaxy is the quiescent galaxy at $z=4.9$ in \cite{deGraaff.etal.2025} and this work. }. That study adopted a redshift-dependent sSFR cut that roughly approximates a rest-frame UVJ selection. \cite{Long.etal.2024} selected quiescent galaxies from the same dataset (CEERS/EGS) based on observed-frame colors, using color cuts informed by model templates. \cite{Valentino.etal.2023} derived number densities by combining the quiescent galaxies selected with rest-frame colors from all available fields at the time. We choose the version of their results computed with the massive sub-sample, which was selected with the padded UVJ selection criteria. 

Most of the number densities derived from photometric samples appear to be ($ 0.3 \sim 0.5 \,\mathrm{dex}$) higher than ours at $3<z<5$. These studies overlap with RUBIES partially in CEERS/EGS or PRIMER/UDS. While the non-overlapping sources between RUBIES and these studies can contribute to this discrepancy, it is also likely that the discrepancy between \cite{Carnall.etal.2023} and RUBIES is driven by the contamination of dusty star-forming objects ($\sim 10\%-35\%$ in rest-frame {\it UVJ} selection, as reported in \cite{Leja.etal.2019b}, \cite{Antwi-Danso.etal.2023}, \cite{Nanayakkara.etal.2025}, or Section \ref{Sec: probe UVJ}). We note that \cite{Long.etal.2024} uses empirical colors to select quiescent galaxies for which the systematic contamination or incompleteness is challenging to estimate. \cite{Valentino.etal.2023} report a slightly lower number density at $3<z<4$, which can be explained by its mass limit being $0.1 \,\mathrm{dex}$ higher than the effective stellar mass limit in our sample at this redshift. In addition, the known over-densities at $3<z<4$ in CEERS/EGS (``Cosmic Vine", \citealp{Jin.etal.2024}) can drive the number densities in this work to be higher, since \cite{Valentino.etal.2023} is further constrained by various other line-of-sights. However, the contribution to the final statistic from intrinsic over-densities in CEERS is likely small, since similar results are obtained in \cite{Baker.etal.2024}.

In the right panel of Figure \ref{fig:number-densities-observations}, we separate the Criterion 1 galaxies in this sample by $\mathrm{t_{90}}$ into young and old quiescent galaxies and calculate their number densities respectively using parent catalog sources in the RUBIES footprint.  This is a key benefit of including spectroscopic information in our analysis. We further compare these to (pre-JWST) literature measurements at lower redshifts that rely on spectroscopy \citep{Belli.etal.2019,Forrest.etal.2018} or photometry \citep{Clausen.etal.2024} to estimate galaxy ages. We find that the number density of older quiescent galaxies from pre-JWST samples is systematically lower than the median estimation from RUBIES at $2<z<3$, although these results are marginally within our measurement uncertainties. However, for young quiescent galaxies, there is a large discrepancy (almost $ 1\, \mathrm{dex}$) in number density at $z>3$ between RUBIES and the only non-JWST result at these redshifts \citep{Forrest.etal.2018}, potentially due to differences in effective mass limits. We find a relatively consistent number density (within $0.3 \,\mathrm{dex}$) with non-JWST literature results with similar minimum mass at $2<z<3$ \citep{Forrest.etal.2018, Clausen.etal.2024}. The young quiescent galaxy number density at $2<z<3$ reported by \cite{Belli.etal.2019} found much lower ($\sim 0.8 \, \mathrm{dex}$) number densities, but for a much more massive sample ($\mathrm{M_{*}> 10^{10.8}M_{\odot}}$ versus $\mathrm{M_{*}> 10^{10.3}M_{\odot}}$ for RUBIES). 

At $z\sim3$, these ground-based samples likely miss both the old and young quiescent galaxies compared to any JWST sample such as RUBIES, due to several factors. The selection of pre-JWST era samples at these epochs, such as the one in \cite{Forrest.etal.2018} \citep[following][]{Kriek.etal.2011}, primarily relied on magnitude-limited K-band surveys. At the low-mass or high redshifts, quiescent galaxies will fall below K-band selection limits. This is especially relevant for the oldest galaxies, which have the highest M/L in the K band. This incompleteness at $z>3$ is further exacerbated by the lack of rest-frame NIR coverage, which requires extrapolation to compute rest-frame V-J colors for common UVJ color selection\citep{Antwi-Danso.etal.2023}. Meanwhile, JWST samples reach deeper magnitude limits, cover longer wavelengths, and have yielded relatively consistent number densities of quiescent galaxies at $z>3$.

Constraining the number density of the first old quiescent galaxies at cosmic noon provides an opportunity to validate the number density of young quiescent galaxies at earlier epochs. After approximately $\mathrm{\sim 0.5 - 1 \, Gyr}$, young quiescent galaxies at a given epoch lose the A-type stars that dominate their SEDs and thus become part of the old quiescent population in the next epoch \citep{Whitaker.etal.2012}. Therefore, the old quiescent population should be a cumulative ensemble of all quiescent galaxies quenched $\mathrm{\sim 0.5 - 1 \, Gyr}$ ago, if rejuvenation or major mergers are infrequent among these galaxies. Since the number density of young quiescent galaxies appears to drop steeply at $4<z<5$ (as previously discussed) and the universe is only $\mathrm{\sim 1 \, Gyr}$ old at $z\sim 5$, we expect the majority of the old quiescent population uncovered at $2<z<3$ to be descendants of the young at $z\sim4$. Our measurements are consistent with this picture: the number density of old quiescent galaxies at $2<z<3$ is indeed similar to or slightly higher than those of the young $\mathrm{\sim 1 \, Gyr}$ beforehand. However, we note that the number density of old quiescent galaxies is fairly uncertain ($\mathrm{\sim 0.5\,dex}$) even with a broad redshift bin (equivalent to $\mathrm{\sim 1 \, Gyr}$). In the future, a larger spectroscopic census of the old quiescent population at $2<z<3$ will help further test this interpretation. 

\subsection{There Are More Massive Quiescent Galaxies in the Early Universe Than Predicted by Simulations}

\begin{figure*}[!htb]
    \centering
    \includegraphics[width = 1.\textwidth]{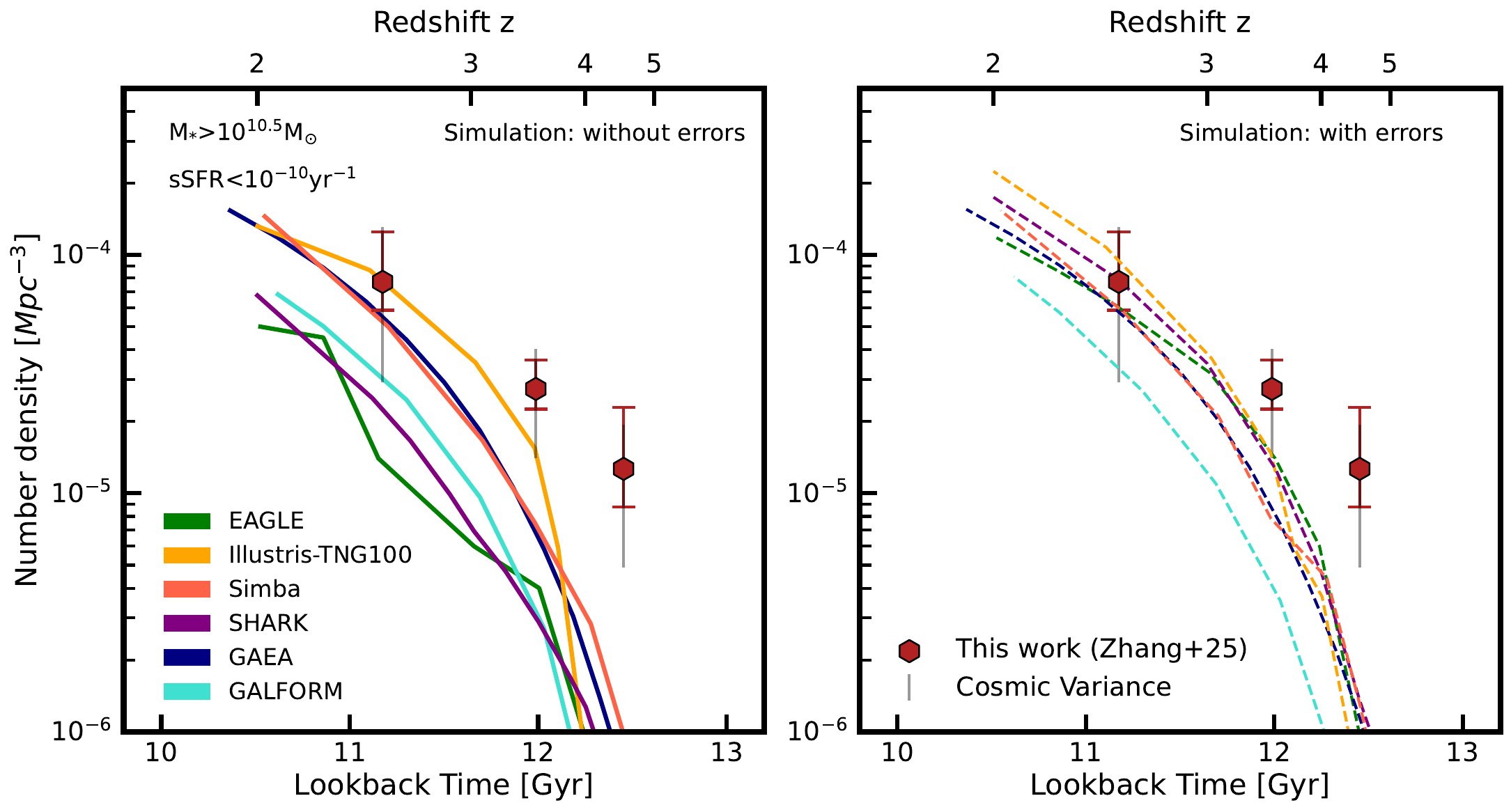}
    \caption{The quiescent galaxy number density reported in this work compared to simulation predictions in the literature. Both the observed and simulation values are computed assuming a quiescence criterion of $\mathrm sSFR < 10^{-10} \, yr^{-1}$ and a mass limit of $\mathrm log(M_{*}/M_{\odot}) >10.5$. Note that we have removed lower mass galaxies ($\mathrm {log}(M_{*}/M_{\odot}) <10.5$) in the $2<z<3$ bin to achieve a uniform mass limit in this comparison.} The simulation values shown in the right panel are computed with random errors in $\mathrm M_{*}$ and $\mathrm {SFR}$ while those in the left are computed without considering those errors. At $z>3$, all simulations shown here under-predict the population abundance of quiescent galaxies.
    \label{fig:number-densities-simulations}
\end{figure*}

In Figure \ref{fig:number-densities-simulations}, we compare the observed number density of quiescent galaxies derived from this sample to various simulation predictions discussed in \cite{Lagos.etal.2025}. Three are semi-analytical models: \textsc{Shark} \citep{Lagos.etal.2018,Lagos.etal.2024}, \textsc{GAEA} \citep{DeLucia.etal.2014,Hirschmann.etal.2016,DeLucia.etal.2024}, and \textsc{Galform} \citep{Lacey.etal.2016}. The other three are cosmological hydrodynamical simulations: \textsc{Eagle} \citep{Crain.etal.2015,Schaye.etal.2015,McAlpine.etal.2016}, \textsc{IllustrisTNG} \citep{Pillepich.etal.2018,Springel.etal.2018}, and \textsc{Simba} \citep{Dave.etal.2019}. All three semi-analytical models correspond to co-moving volumes of $\mathrm{\sim (700\,cMpc)^3}$, and the three cosmological hydrodynamical simulations have co-moving box sizes of $\mathrm{\sim (100\,cMpc)^3-(150\,cMpc)^3}$. All simulation predictions shown here assume a quiescence criterion of $\mathrm{sSFR} < 10^{-10} \, yr^{-1}$ and a stellar mass limit of $\mathrm log(M_{*}/M_{\odot}) >10.5$, regardless of dust attenuation. The simulation number densities shown in the left panel select galaxies by their exact values in stellar mass and SFR, while those in the right panel additionally consider the random errors in these properties, as described in \cite{Lagos.etal.2025} or \cite{DeLucia.etal.2024}. For the curves shown in the right panel, number densities are recalculated after convolving the galaxy mass function or SFR distribution in these simulations with a Gaussian to mimic the scattering of the galaxy population in mass or SFR due to errors. We assume errors in stellar mass and SFR to be independent and these Gaussians are centered at 0 with widths of $0.3\, \mathrm{dex}$ (except for \textsc{GAEA}, where the widths are $0.35\, \mathrm{dex}$). These choices of Gaussian widths for error convolution are motivated by the typical uncertainties in stellar mass and SFR inferred from multi-wavelength observations \citep{Robotham.etal.2020,Bellstedt.etal.2025}. Notably, incorporating scatter to emulate the effects of measurement uncertainties systematically increases the number densities in \textsc{Eagle} and \textsc{Shark}. This is likely because $>10^{10.5}$ M$_{\odot}$ falls in the exponential decline of the quiescent galaxy stellar mass functions in these simulations \citep{Lagos.etal.2025}, introducing a net upward Eddington bias and inflating number densities. For the observed quiescent galaxy number density shown in these panels, we take the fiducial sample (Criterion 1) and remove galaxies with $\mathrm{\log(M_{*}/M_{\odot}) <10.5}$ in the $2<z<3$ bin, in order to be consistent with the sSFR or mass limit in these simulations. \cite{Baker.etal.2025} has shown that the observed quiescent galaxy stellar mass function is flat at $\mathrm{\log(M_{*}/M_{\odot}) \sim 10.5}$ cutoff at $2<z<4$. Therefore, the number density derived from RUBIES is potentially not sensitive to the Eddington bias discussed above at $z<4$.

At face value, only \textsc{Simba}, \textsc{IllustrisTNG}, and \textsc{GAEA} agree well with the observed number density of quiescent galaxies at $2<z<3$. After emulating measurement uncertainties, all six simulations are largely consistent with observational data at cosmic noon. However, at earlier times, all simulations under-predict the observed number densities, by $ \sim 0.4 \, \mathrm{dex}$ at $3<z<4$ and $\mathrm \sim 1 \, \mathrm{dex}$ at $4<z<5$. This discrepancy persists regardless of our empirical definition of quiescence; for example, using the evolving quiescence criterion of $\mathrm{sSFR} < 0.2/t_{universe}(z) \, \mathrm{Gyr}^{-1}$ produces similar results. Given the median theoretical number density, we would expect to find $\sim 1$ quiescent galaxy at $\mathrm{\log(M_{*}/M_{\odot}) >10.5}$ at $4<z<5$ in the total sky area covered by EGS and UDS ($\mathrm{304\, arcmin^2}$). Yet we find five such galaxies in the $\sim50\%$ covered by the RUBIES survey ($\mathrm{150\, arcmin^2}$). We note that cosmic variance plagues simulations and observations alike. These cosmological simulations have small box sizes ($\mathrm{\sim (100\,cMpc)^3}$), for which a number density of $\mathrm{\sim 10^{-6}\,cMpc^{-3}}$ corresponds to one object in the entire simulation box. At $4<z<5$, their number densities are sensitive to random counting error. However, we emphasize that this comparison is still meaningful since the number density of quiescent galaxies in the RUBIES sample is an order of magnitude higher ( $\mathrm{\sim 10^{-5}\,cMpc^{-3}}$). Given our consistency with previous studies, we conclude that the dramatic discrepancy between observed and predicted quiescent galaxy populations before $z>3$ is unlikely to be attributed to the contamination within photometric samples or targeting biases in spectroscopic studies.

Although we expect the effect to be small, our measured number densities could be slightly underestimated at the highest redshifts due to the effective magnitude completeness limits within the RUBIES survey. The magnitude limit ($\mathrm{F444W <24}$) of our fiducial parent sample corresponds to stellar mass limits of $\mathrm{10^{10.3} M_{\odot}}$ ($2<z<3$), $\mathrm{10^{10.5} M_{\odot}}$ ($3<z<4$), and $\mathrm{10^{10.6} M_{\odot}}$ ($4<z<5$), as discussed in Appendix \ref{appendix: effective limit}. Therefore, RUBIES could have failed to target galaxies $\mathrm{10^{10.5} <\log M_*< 10^{10.6}M_{\odot}}$ with high intrinsic mass-to-light ratios or near $z\sim5$. We expect this effect to be insignificant given our simulations (Figure \ref{fig:effective mass limit}). We note that it is further possible, though unlikely, that this sample is missing a significant population of heavily dust-attenuated $\mathrm{A_v >0.7}$ quiescent galaxies (see details in Appendix \ref{appendix: effective limit}). Although most quiescent galaxies at $z<2.5$ have $\mathrm{A_v <0.75}$ \citep[e.g.,][]{Suess.etal.2019a, Siegel.etal.2025}, rare counter examples with significant dust-reddening in the core ($\mathrm{A_v >0.75}$, \citealp{Setton.etal.2024a,Siegel.etal.2025}) or the outskirts \citep{Ji.etal.2024} exist. However, these sources of sample incompleteness would only exaggerate the tension between simulations and observations.

Among these simulations and models, the wide range of predicted quiescent number densities at a given redshift is mainly due to the different implementations of AGN feedback (see \citealt{Lagos.etal.2025} for detailed discussion). Further modifications to the AGN feedback implementation could resolve the current tension in quiescent galaxy number densities. Although hard to pinpoint observationally, many lines of evidence point toward the simultaneity of quenching and AGN activity. For example, the AGN incidence rate of massive quiescent galaxies at similar epochs ($\mathrm{\sim 50\%}$ from a multi-wavelength search by \citealp{Baker.etal.2024}; $\mathrm{\sim 20\%}$ from analysis of optical emission lines in \citealp{Martinez-Marin.2024} or \citealp{Ito.etal.2025b}) is much higher than that of the youngest quiescent galaxies at low redshifts \citep{Greene.etal.2020}. A number of the quiescent galaxies in this sample exhibit strong nebular emission lines, which hints at the incidence of nuclear activities, although we defer that analysis to a future study. Including re-ionization physics has been shown to significantly boost the number of quiescent galaxies at earlier times ($z\sim5.5$; \citealp{Chittenden.etal.2025}), which could ultimately become important in resolving tension with future theoretical models.

In the future, the lack of massive quiescent galaxies in state-of-the-art galaxy formation simulations needs to be investigated further to separate two compounding issues. One is the potential overall lack of massive galaxies \citep[either star-forming or quenched; e.g.,][]{Weibel.etal.2024b,Shuntov.etal.2025}, which would point to star formation not being efficient enough or, conversely, feedback being too strong in regulating star formation in the early universe. The second one is AGN feedback itself, and whether the processes it encompasses (e.g., mechanical, radiative, or energetic feedback) are sufficient to quench massive galaxies in the early universe. It is clear that this field is nascent, and further observations of massive quiescent galaxies and their stellar mass distribution over larger samples would provide invaluable constraints for galaxy formation models.

A final resolution is empirical: if most apparently quiescent galaxies host heavily dust-obscured star formation, then the apparent tension could disappear. Testing this would require additional observations of apparently quiescent galaxies at $3<z<5$ in the mid-IR or Far-IR (FIR). All of the spectroscopic identifications of quiescent galaxies thus far rely on interpreting their rest-frame optical-NIR emission. These inferences cannot yet rule out extreme birth-cloud dust attenuation ($\mathrm{A_V \sim 5}$) that could hide instantaneous star formation. This scenario has already been discovered in some $z<1$ optically selected post-starburst galaxies, which have FIR SFR $\sim 1-2 \, \mathrm{dex}$ higher than those inferred from their optical information \citep{Baron.etal.2023}. As revealed by our PCA analysis (also in \citealp{Cooper.etal.2025}), the prevalence of galaxies that simultaneously host evolved stellar populations while being dust-attenuated suggests this is plausibly a more common scenario at $z>3$. If $\sim 90\%$ of the rest-frame-optically-selected quiescent galaxies at $z>4$ turned out to host star formation embedded in optically thick dust, the $\mathrm{1\,dex}$ discrepancy would disappear. Stellar population synthesis modeling of a truly panchromatic sample of quiescent galaxies at $z>3$ could lay this uncertainty to rest; the attenuated radiation from instantaneous star formation would inevitably re-radiate at MIR and FIR, testable by deeper-than-existing observations with facilities such as the Atacama Large Millimeter/submillimeter Array. 

\section{Summary} \label{Sec: Summary}
In this paper, we presented a sample of 17 (Criterion 1) or 20 (Criterion 1+2) spectroscopically-confirmed massive ($ \mathrm{ log (M_{*}/M_{\odot}) > 10.3} $) quiescent galaxies at $2<z<5$ and their physical properties, using JWST NIRSpec PRISM spectra from the RUBIES sample. We developed an efficient methodology to identify quiescent galaxies, performing PCA on all public DJA PRISM spectra to establish eigenspectra and identify quiescent galaxy candidates in RUBIES. We infer the properties of the stellar populations by modeling the NIRSpec PRISM spectra and NIRCam photometry for each candidate with \texttt{Prospector} and converge on a final spectroscopic sample. We leverage the well-defined color-magnitude targeting strategy of the RUBIES survey to derive the number density of young, old, and total quiescent galaxies between $3<z<5$. We have obtained the following findings:

\begin{itemize}

\item{We compare our spectroscopic sample of quiescent galaxies to photometric rest-frame color selection methods, such as {\it UVJ} and {\it $u_s$$g_s$$i_s$}. We estimate that such selections will be significantly contaminated ($\sim 35\%$ and $\sim 60\%$, respectively), even without uncertainties due to photometric redshifts and/or extrapolation due to e.g., NIRCam coverage at $z\gtrsim3$ \citep{Antwi-Danso.etal.2023}.}

\item{We find that the number densities of both young and old quiescent galaxies in our spectroscopic sample are systematically higher than pre-JWST samples above $z>2$, but consistent with other JWST studies. Although only found at cosmic noon, the number density of older quiescent galaxies at $2<z<3$ is consistent with the expected aging population from the previous $\mathrm{\sim 1 \, Gyr}$.}

\item{As reported in previous studies, we find that massive quiescent galaxies at $z>3$ are much more common than predictions from six state-of-the-art cosmological galaxy formation simulations. This discrepancy at $z>4$ is unambiguous even when the cosmic variance is included, as the number density of massive quiescent galaxies estimated with the RUBIES sample is 10 times greater than the simulation prediction at this epoch.}

\end{itemize}

Understanding the formation and quenching of the first massive quiescent galaxies, indeed even just matching number densities, will require efforts on the theoretical and observational fronts. For this rare population, beating down the uncertainties due to cosmic variance by increasing surveyed volumes is critical. The co-moving volume probed by RUBIES in each one of the redshift bins is merely $\mathrm{\sim 5\cdot10^5 \, cMpc^3}$, compared to simulation volumes that are typically $\mathrm{\sim 10^6 - 10^8 \, cMpc^3}$. Dramatically increasing the area of the sky probed by JWST imaging would be an obvious first step, although we emphasize the high contamination rates of quiescent samples even with CEERS/PRIMER NIRCam photometric coverage and spectroscopic redshifts. This would be much worse in shallower imaging and/or with sparsely sampled SEDs from e.g., COSMOS-Web. JWST parallel imaging surveys, such as PANORAMIC \citep{Williams.etal.2025}, can provide an opportunity to efficiently cover large areas (with many filters) and provide independent fields that optimally minimize cosmic variance uncertainties \citep{Jespersen.etal.2025}. However, spectroscopic confirmation will always be necessary, ideally leveraging larger imaging surveys for targeting using well-characterized selection functions as in RUBIES. Ideally, these samples will comprise maximal multi-wavelength data, including coverage in the MIR and FIR, to conclusively confirm quiescence. Finally, even at cosmic noon, the number densities and ages of the descendants of these extreme, early quiescent galaxies are poorly constrained. Thus, wide-area large spectroscopic surveys like the Prime Focus Spectrograph Survey \citep{Greene.etal.2022} or MOONS \citep{Maiolino.etal.2020} promise to provide interesting insights into the number densities and star formation histories of old quiescent systems at cosmic noon and indirectly test their earliest histories.

\section{Acknowledgments}
We thank Alan Pearl for his contribution to estimating the cosmic variance contribution to the number density uncertainties in this work. We thank Hans-Walter Rix for the valuable discussion on the selection function and correcting the spectroscopic incompleteness of massive quiescent galaxies in RUBIES.

Some of the data products presented herein were retrieved from the Dawn JWST Archive (DJA). DJA is an initiative of the Cosmic Dawn Center (DAWN), which is funded by the Danish National Research Foundation under grant DNRF140.

The Cosmic Dawn Center is funded by the Danish National Research Foundation (DNRF) under grant \#140. 

Support for this work was provided by The Brinson Foundation through a Brinson Prize Fellowship grant. 

RB gratefully acknowledges support from the Research Corporation for Scientific Advancement (RCSA) Cottrell Scholar Award ID No: 27587.

The work of CCW is supported by NOIRLab, which is managed by the Association of Universities for Research in Astronomy (AURA) under a cooperative agreement with the National Science Foundation.

TBM was supported by a CIERA Postdoctoral Fellowship.

This work is based in part on observations made with the NASA/ESA/CSA James Webb Space Telescope. The data were obtained from the Mikulski Archive for Space Telescopes at the Space Telescope Science Institute, which is operated by the Association of Universities for Research in Astronomy, Inc., under NASA contract NAS 5-03127 for JWST. These observations are associated with program numbers 1180, 1181, 1210, 1211, 1213, 1214, 1215, 1286, 1345, 1433, 1747, 2028, 2073, 2198, 2282, 2561, 2565, 2750, 2756, 2767, 3073, 3215, 4233, 4446, 4557, 6368, 6541, and 6585. The specific observations analyzed can be accessed via \dataset[doi: 10.17909/sjsj-8p46]{https://doi.org/10.17909/sjsj-8p46}.

Support for program no. 4233 was provided by NASA through a grant from the Space Telescope Science Institute, which is operated by the Association of Universities for Research in Astronomy, Inc., under NASA contract NAS 5-03127.

The authors acknowledge the CEERS and PRIMER teams for developing their observing program with a zero-exclusive access period.

\facilities{ 
JWST - James Webb Space Telescope
}

\software{Astropy \citep{astropy:2013,astropy:2018,astropy:2022}, Scipy \citep{2020SciPy}, Photutils \citep{photutils}, SpectRes \citep{Carnall.etal.2017}, scikit-learn \citep{scikit-learn}, Prospector\citep{Leja.etal.2017,Johnson.Leja.2017,Johnson.2021}}

\bibliography{sample631}{}
\bibliographystyle{aasjournal}

\appendix 

\section{The SFHs and SEDs of the remaining PCA selected galaxies}\label{appendix: all_gallery}

In Figure \ref{fig:all_gallery01} and \ref{fig:all_gallery02}, we present the best-fitting SFHs and model SEDs, along with the observed NIRCam photometry and NIRSpec PRISM spectra, of the remaining 18 RUBIES massive quiescent galaxies ($\mathrm{sSFR_{16th} <10^{-10} \,yr^{-1}}$) and 9 RUBIES dusty impostors with similar SCs to the quiescent galaxies. The properties of the full RUBIES quiescent sample are tabulated in Table \ref{tab:all_measurements}.

\begin{figure*}[h!tb]
    \centering
    \includegraphics[width = 0.97\textwidth]{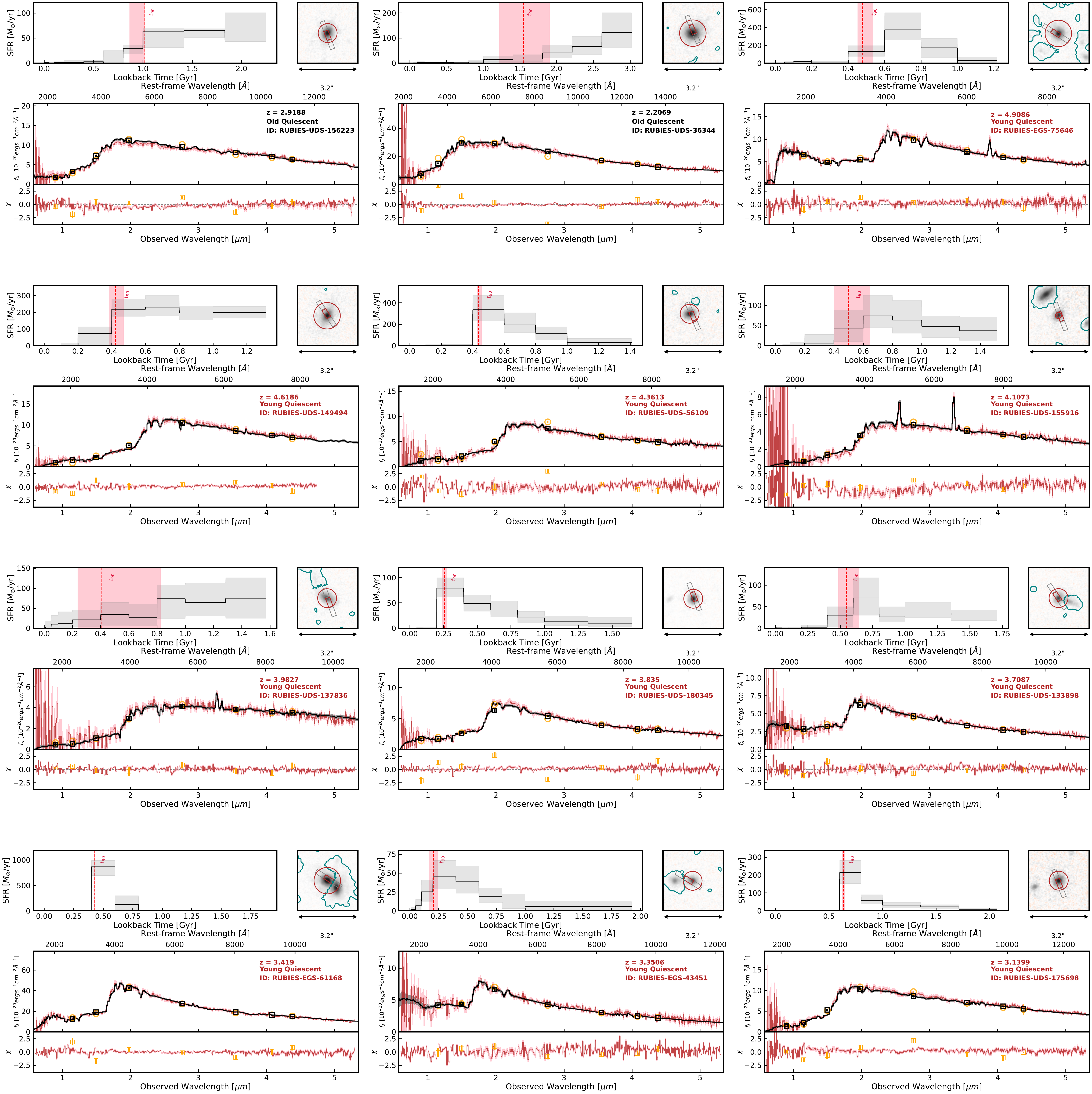}
    \caption{A gallery of the observed SEDs, best-fitting models, and SFHs of all the remaining quiescent galaxies in our sample as well as a few selected unquenched impostors. The plotting convention follows Figure \ref{fig:SED_young_QG}. In addition, the teal contour in the image postage represents the mask image used during the aperture photometry extraction.} 
    \label{fig:all_gallery01}
\end{figure*}

\begin{figure*}[h!tb]
    \centering
    \includegraphics[width = 0.97\textwidth]{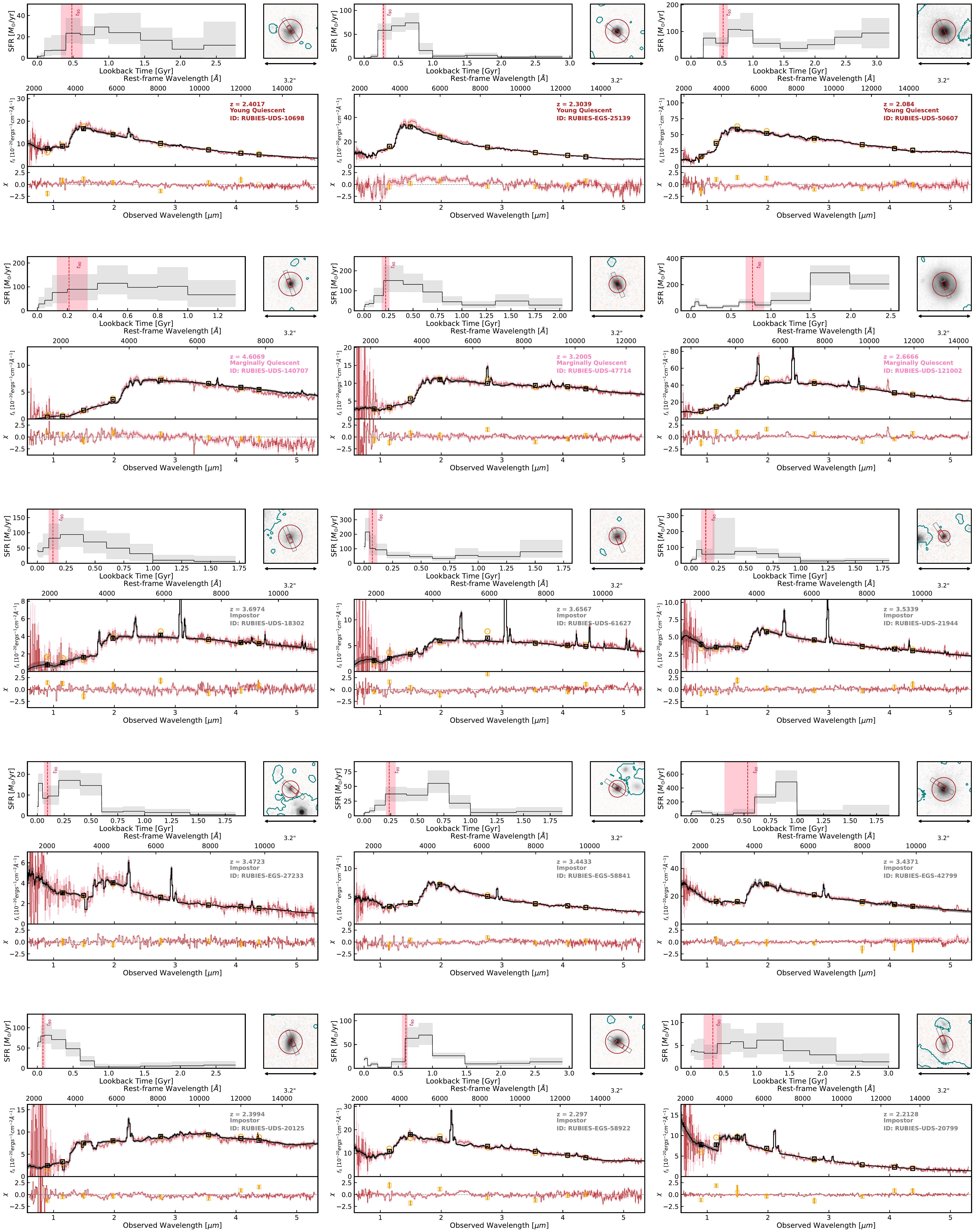}
    \caption{Continued. The plotting convention follows Figure \ref{fig:all_gallery01}.}
    \label{fig:all_gallery02}
\end{figure*}

\movetabledown=1in
\begin{table*}[h]
\begin{rotatetable*}
\begin{center}
\caption{RUBIES Massive Quiescent Galaxies}
    
    \begin{tabular}{ccccccccccccc}
    
    \hline
    \hline
      Category & ID & $z_{spec}$ & RA & Dec& $\mathrm{log(M_{*})}$ & $\mathrm{log(sSFR)}$ & $\mathrm{t_{90}}$ & SC$_0$ &SC$_1$&SC$_2$&SC$_3$  \\
      & & & & &[$\mathrm{M_{\odot}}$]& $[\mathrm{yr^{-1}}]$ & $\mathrm{Gyr}$ & & & & \\
     \hline
     Old Quiescent & RUBIES-UDS-156223 &2.9188 &34.334606 &-5.127057& $10.72^{+0.04}_{-0.01}$&
      $-10.71^{+0.15}_{-0.12} $ & $1.01^{+0.01}_{-0.15}$ &-1.44 & 0.16& -1.00 & 0.29\\
     Old Quiescent & RUBIES-EGS-42328 &2.6873& 214.978556 &52.921542& $10.48^{+0.05}_{-0.04}$&
      $-10.9^{+0.44}_{-1.12} $ & $1.11^{+0.13}_{-0.12}$&-1.52 & 0.44 &-1.09 & 0.53\\
     Old Quiescent &RUBIES-UDS-36344 &2.2069 &34.269542 &-5.252727& $10.86^{+0.03}_{-0.05}$&
      $-11.02^{+0.4}_{-0.74} $ & $1.55^{+0.36}_{-0.33}$&-1.49 & 0.32 &-1.14 & 0.30\\
     Young Quiescent &RUBIES-EGS-75646\footnote{Also in \cite{deGraaff.etal.2025}}&4.9086 &214.915546 &52.949018& $11.02^{+0.04}_{-0.05}$&
      $-11.68^{+0.6}_{-1.76} $ & $0.48^{+0.06}_{-0.03}$&-1.14 &-0.36 &-0.69 &-0.09\\
      Young Quiescent &RUBIES-UDS-149494\footnote{Also in \cite{Carnall.etal.2024}}&4.6186& 34.399676& -5.136348& $11.15^{+0.05}_{-0.04}$&
      $-11.91^{+0.83}_{-1.74} $ & $0.42^{+0.05}_{-0.04}$&-1.46& -0.14& -0.93 & 0.11\\
      Young Quiescent &RUBIES-UDS-56109& 4.3613& 34.280515& -5.217214& $11.02^{+0.06}_{-0.06}$&
      $-11.37^{+0.58}_{-1.33} $ & $0.44^{+0.02}_{-0.01}$&-1.50 & 0.00 &-1.03 & 0.10\\
      Young Quiescent &RUBIES-UDS-155916 &4.1073 &34.317031 &-5.127611& $10.65^{+0.04}_{-0.03}$&
      $-11.13^{+0.7}_{-1.42} $ & $0.50^{+0.15}_{-0.10}$&-1.55 & 0.19 &-1.05  &0.29\\
      Young Quiescent &RUBIES-UDS-12594 &3.9874 &34.368535 &-5.299475& $10.97^{+0.05}_{-0.05}$&
      $-11.43^{+0.79}_{-2.00} $ & $0.49^{+0.16}_{-0.04}$&-1.47 & 0.10 &-0.91 & 0.14\\
      Young Quiescent &RUBIES-UDS-137836 &3.9827&34.360159 &-5.153092& $10.74^{+0.06}_{-0.06}$&
      $-10.75^{+0.56}_{-1.22} $ & $0.41^{+0.42}_{-0.17}$&-1.45 & 0.22 &-1.09 & 0.25\\
      Young Quiescent &RUBIES-UDS-180345 &3.835 &34.209839 &-5.091602& $10.51^{+0.03}_{-0.03}$&
      $-11.19^{+0.62}_{-1.58} $ & $0.26^{+0.02}_{-0.01}$&-1.40 &-0.37 &-0.89 &-0.13\\
      Young Quiescent &RUBIES-UDS-133898 &3.7087 &34.485164 &-5.157813& $10.54^{+0.05}_{-0.04}$&
      $-10.49^{+0.43}_{-1.11} $ & $0.55^{+0.10}_{-0.06}$&-1.31 &-0.16 &-0.80 &-0.04\\
      Young Quiescent &RUBIES-EGS-61168 \footnote{Also in \cite{Ito.etal.2025} and \cite{Nanayakkara.etal.2025}} &3.419 &214.866053 &52.884257& $11.15^{+0.03}_{-0.03}$&
      $-14.19^{+1.62}_{-3.6} $ & $0.42^{+0.01}_{-0.00}$&-1.35 &-0.39 &-0.86 &-0.11\\
      Young Quiescent &RUBIES-EGS-43451 &3.3506 &214.909553 &52.875028& $10.35^{+0.07}_{-0.07}$&
      $-10.48^{+0.45}_{-0.92} $ & $0.21^{+0.03}_{-0.04}$&-1.05 &-0.28 &-0.68 &-0.03\\
      Young Quiescent &RUBIES-UDS-175698 &3.1399 &34.227642 &-5.099286& $10.73^{+0.03}_{-0.04}$&
      $-12.03^{+0.91}_{-2.31} $ & $0.64^{+0.01}_{-0.01}$&-1.52 & 0.06 &-1.02 & 0.23\\
      Young Quiescent &RUBIES-UDS-10698 &2.4017 &34.406462 &-5.302878& $10.49^{+0.07}_{-0.07}$&
      $-10.69^{+0.37}_{-0.85} $ & $0.48^{+0.15}_{-0.15}$&-1.11 &-0.09 &-0.74  &0.01\\
      Young Quiescent &RUBIES-EGS-25139 &2.3039 &215.113914 &52.977721& $10.57^{+0.02}_{-0.03}$&
      $-10.97^{+0.43}_{-0.88} $ & $0.28^{+0.04}_{-0.02}$&-1.41 &-0.25 &-0.95 &-0.06\\
      Young Quiescent &RUBIES-UDS-50607 &2.084 &34.277979 &-5.227332& $11.15^{+0.04}_{-0.05}$&
      $-11.44^{+0.65}_{-1.84} $ & $0.52^{+0.07}_{-0.06}$&-1.47 & 0.13 &-0.99  &0.20\\
      Marginally Quiescent &RUBIES-UDS-140707 &4.6069 &34.365084 &-5.148848& $10.92^{+0.06}_{-0.05}$&
      $-9.84^{+0.26}_{-0.40} $ & $0.21^{+0.12}_{-0.08}$&-1.41 &-0.09 &-0.81  &0.12\\
      Marginally Quiescent &RUBIES-UDS-47714 &3.2005 &34.25891 &-5.232334& $10.99^{+0.06}_{-0.06}$&
      $-9.71^{+0.20}_{-0.36} $ & $0.22^{+0.04}_{-0.04}$&-1.48  &0.33 &-0.97  &0.31\\
      Marginally Quiescent &RUBIES-UDS-121002 &2.6666 &34.30467 &-5.175739& $11.33^{+0.03}_{-0.02}$&
      $-9.99^{+0.09}_{-0.10} $ & $0.77^{+0.14}_{-0.08}$&-1.41 & 0.49 &-0.99 & 0.47\\
      
      \hline
    \end{tabular}
    {\footnotesize \tablecomments{This table will be available in a machine-readable form.}}
    \label{tab:all_measurements}
\end{center}
\end{rotatetable*}
\end{table*}

\section{Determining the effective limits of this sample in $\mathrm{M_{*}}$ and $A_v$}\label{appendix: effective limit}

\begin{figure*}[h!tb]
    \centering
    \includegraphics[width = 0.97\textwidth]{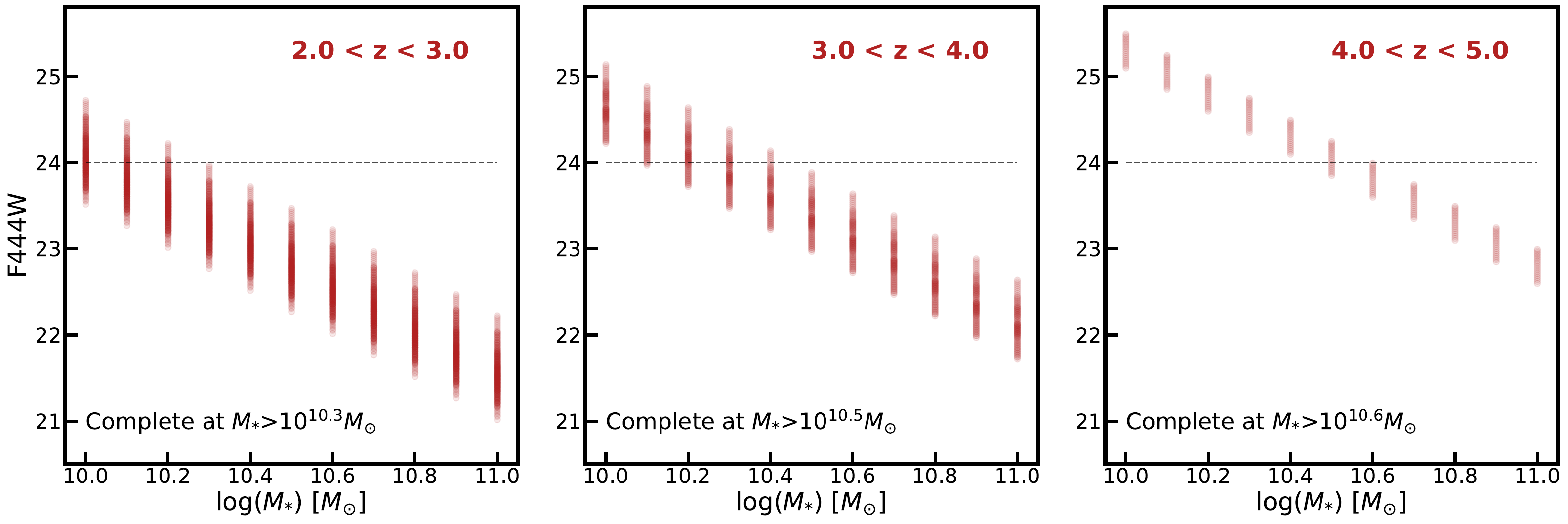}
    \caption{The F444W magnitudes of quiescent galaxies in each redshift bin predicted from the best-fitting Prospector models of quiescent galaxy ($\mathrm sSFR < 10^{-10} yr^{-1}$) in this sample, using a grid of stellar mass and redshift. The magnitude limit of this sample ($\mathrm m_{F444W}<24$) would have included any quiescent galaxies with $\mathrm log(M_{*}/M_{\odot})> 10.3 $ at $2<z<3$, $\mathrm log(M_{*}/M_{\odot})> 10.5 $ at $3<z<4$, and $\mathrm log(M_{*}/M_{\odot})> 10.6 $ at $4<z<5$, assuming the mass-to-light ratios of quiescent galaxies in this sample are representative of the entire quiescent population in each epoch.} 
    \label{fig:effective mass limit}
\end{figure*}

We divide the final quiescent sample (Criterion 1) into three redshift bins ($[2,3], [3,4], $ and $[4,5]$). For galaxies in each redshift bin, we obtain an ensemble of their analogs by taking their best-fitting \texttt{Prospector} models, redshifting their model SEDs to a grid of redshifts within the corresponding bin interval, and rescaling these model SEDs to a grid of stellar masses within $\mathrm{10^{10}M_{\odot}<logM_{*} < 10^{11}M_{\odot}}$. We note that these analog SEDs include both the cosmic dimming due to their redshifts and the intrinsic brightness due to their stellar masses. Therefore, we expect these analog SEDs to approximately resemble those of the quiescent population that would have been observed at these redshifts and masses. We derive the corresponding F444W magnitudes of these dimmed and redshifted analogs, which are shown in Figure \ref{fig:effective mass limit}. Using these predicted F444W magnitudes, we determine the stellar mass at which all analogs in each redshift bin would be brighter than our magnitude limit. Our magnitude-limited selection would have been complete above $\mathrm{10^{10.3}M_{\odot}}$ at $2<z<3$, above $\mathrm{10^{10.5}M_{\odot}}$ at $3<z<4$, and above $\mathrm{10^{10.6}M_{\odot}}$ at $4<z<5$. 

\begin{figure*}[h!tb]
    \centering
    \includegraphics[width = 0.6\textwidth]{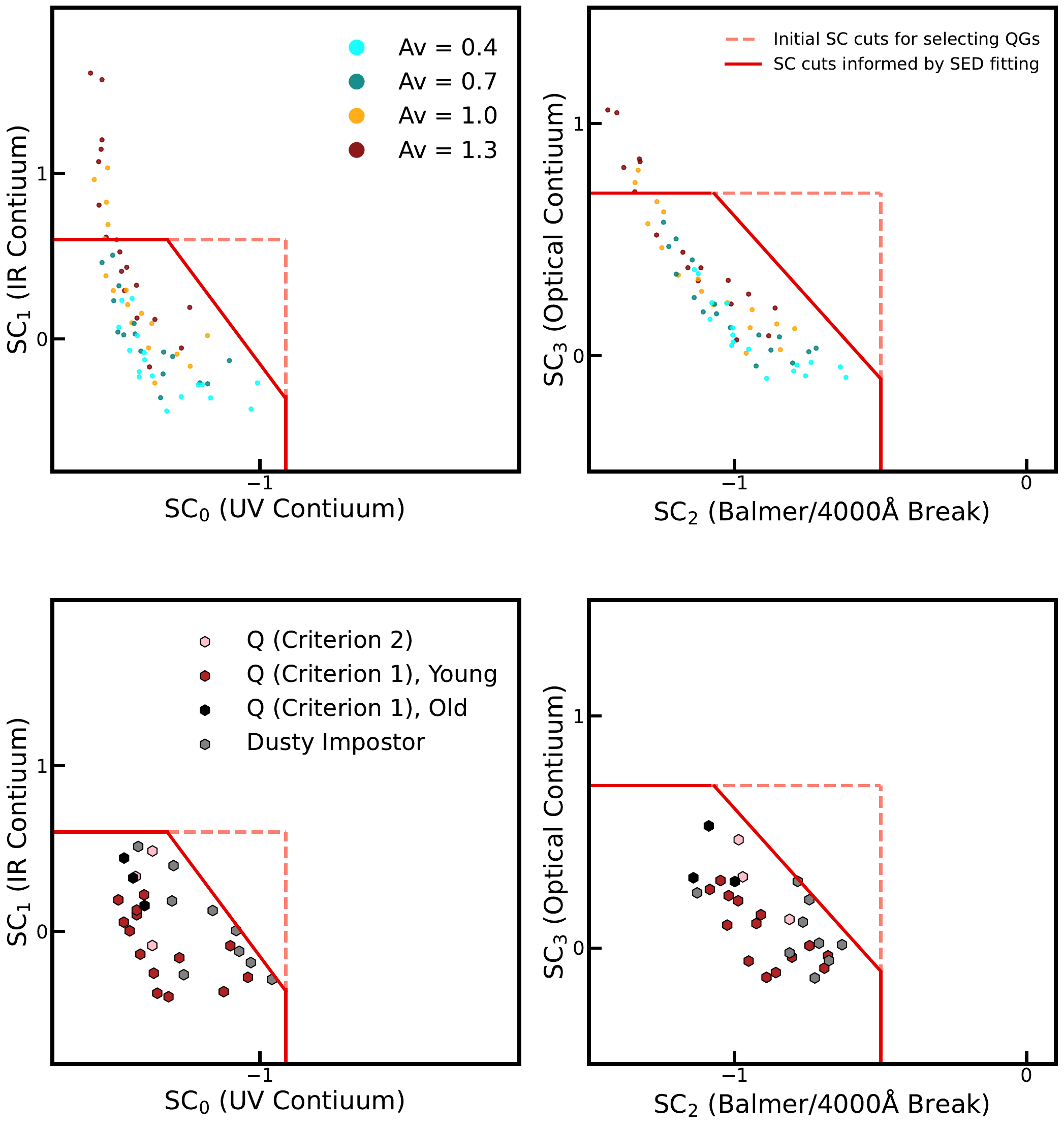}
    \caption{Top row: The SCs of quiescent galaxies predicted from the best-fitting Prospector models of quiescent galaxy ($\mathrm sSFR < 10^{-10} yr^{-1}$) in this sample, using a grid of dust attenuation ($A_V$). The SC cuts adopted by this sample selection would have included all of the quiescent galaxies for $A_V <0.7$. Bottom row: The SCs of RUBIES massive quiescent galaxies and dusty impostors selected by these SC cuts.} 
    \label{fig:effective av limit}
\end{figure*}

In order to determine how the SCs of quiescent galaxies depend on dust attenuation, we take the best-fitting \texttt{Prospector} models of the fiducial quiescent sample (Criterion 1; $\mathrm{sSFR_{50th} <10^{-10} \,yr^{-1}}$) and generate model spectra for their analogs of different dust attenuation levels, using a grid of $\mathrm{A_v}$ parameter inputs. For each analog model spectrum, the model setup remains the same and all other model parameters are fixed to the best-fitting values. Following the same procedure described in Section \ref{sec:pca_selection}, we de-redshift and resample the model spectra to the same wavelength grid described with \texttt{SpectRes}. These resampled analog model spectra are also normalized by flux density at rest-frame $\mathrm{4500 \AA}$. To compute the four SCs, we linearly solve for the four coefficients of eigenspectra to minimize chi-squared, using a standard package in \texttt{Scipy}. The SCs of the fiducial quiescent sample at four selected dust attenuations are shown in the top panels Figure \ref{fig:effective av limit}. Overall, as $A_V$ increases, $\mathrm{SC_{1}}$ and $\mathrm{SC_{3}}$ of these galaxies increase while $\mathrm{SC_{0}}$ and $\mathrm{SC_{2}}$ decrease. In addition to the initial SC cuts, we adopt SC cuts in $\mathrm{SC_{0} -SC_{1}}$ and $\mathrm{SC_{2} -SC_{3}}$ that are parallel to these trends in SC as $\mathrm{A_V}$ increases, eliminating the SC regions that are not occupied by any quiescent galaxies at any $\mathrm{A_V}$. The fiducial quiescent sample in this work could have been fully selected by the refined SC cuts for $\mathrm{A_V<0.7}$. 

A final caveat regarding these SC selections is that a given source can shift $\sim 0.1$ or less in these SC spaces, due to nuances in the spectral shape when adopting different flux calibration. Tracking these systematic uncertainties in SCs is challenging because the flux calibration for each source is unique and complicated. To prevent an under-selection of quiescent galaxies due to these uncertainties in SCs, the final SC cuts we have adopted are still considerably generous, and we have reserved space between these SC cuts and the quiescent galaxies confirmed in this study (see bottom panels of Figure \ref{fig:effective av limit}).

\end{document}